\documentclass[preprint,aps,nofootinbib]{revtex4}

\newcommand{\be}{\begin{eqnarray}}
\newcommand{\ee}{\end{eqnarray}}
\newcommand{\non}{\nonumber\\}
\newcommand{\ave}[1]{\left\langle #1 \right\rangle}
\newcommand{\GeV}{\hbox{\,GeV}}
\newcommand{\tw}{\textwidth}
\newcommand{\Ymax}{y_{\rm max}}
\newcommand{\ymax}{{y_{\rm max}}}
\newcommand{\chip}{{\chi_{p}}}
\newcommand{\chif}{{\chi_{f}}}
\newcommand{\etap}{{\eta_{p}}}
\newcommand{\etaf}{{\eta_{f}}}
\newcommand{\sqrts}{\sqrt{s}}

\usepackage{epsf}

\begin{document}

\title{\bf\Large
Universal Transition Curve\\
in Pseudo-Rapidity Distribution
}

\author{\large Sangyong Jeon}
\affiliation{
McGill University, 3600 University Street Montreal, QC H3A-2T8,
Canada}
\affiliation{
RIKEN-BNL Research Center,
Broohaven National Laboratory,
Upton, NY 11973, USA}
\author{\large Vasile Topor Pop}
\affiliation{
McGill University, 3600 University Street Montreal, QC H3A-2T8,
Canada}
\author{\large Marcus Bleicher}
\affiliation{Institut f\"ur Theoretische Physik
J. W. Goethe-Universit\"at
Robert-Mayer-Str. 8-10
60054 Frankfurt am Main, Germany}

\begin{abstract}

 We show that an unambiguous way of determining the universal
 limiting fragmentation
 region is to consider the derivative ($d^2n/d\eta^2$) of the pseudo-rapidity
 distribution per participant pair. 
 In addition, we find that the transition region between the
 fragmentation and the central plateau regions exhibits a second kind of
 universal behavior that is only apparent in $d^2n/d\eta^2$. 
 The $\sqrts$ dependence of the 
 height of the central plateau $(dn/d\eta)_{\eta=0}$ and the total
 charged particle multiplicity $n_{\rm total}$ critically depend on the
 behavior of this universal transition curve.
 Analyzing available RHIC data, we show that $(dn/d\eta)_{\eta=0}$ can be 
 bounded by $\ln^2 s$ and $n_{\rm total}$ can be bounded by $\ln^3 s$.
 We also show that the deuteron-gold data from RHIC has the exactly
 same features as the gold-gold data indicating that these universal
 behaviors are a feature of the initial state parton-nucleus interactions
 and not a consequence of final state interactions.
 Predictions for LHC energy are also given.
\end{abstract}

\maketitle

\section{Introduction}

 Recently, PHOBOS and BRAHMS collaborations at RHIC 
 published a set of intriguing data.  
 Among them are the striking feature of the limiting fragmentation
 \cite{Back:2002wb,Bearden:2001qq}.
 It is reported by PHOBOS that the shifted (by the beam rapidity $\ymax$)
 pseudo-rapidity distribution per
 participant pair ${dn/d\eta} \equiv (dN_{\rm ch}/d\eta)/(N_{\rm part}/2)$ 
 is independent of colliding energy up to 85-90\,\% of the plateau height
 \cite{Back:2002wb}.
 
 If taken at face value, this would imply that the height of the plateau 
 and its $\sqrts$ dependence
 would be almost fully determined by the limiting fragmentation curve and 
 the location of the beginning of the plateau. 
 Also, since the total multiplicity is simply the area under $dn/d\eta$,
 its $\sqrts$ dependence would be largely determined by
 the limiting fragmentation curve as well.

 It is not easy to determine the validity of above statements when only the
 rapidity distributions ($dn/d\eta$) are compared. 
 In this paper we argue that comparing the 
 {\em slopes} ($d^2n/d\eta^2$) is a much better way to determine
 various regions of the rapidity distribution. 
 
 A surprising feature is the existence of another `universal' behavior,
 which is only apparent in the slopes.
 It turns out that $d^2n/d\eta^2$ in the transition region between the
 limiting fragmentation and the central plateau also follows a
 universal curve.  This is not an extension of the limiting fragmentation
 curve.  To our knowledge, this is the first time the existence of this
 second universal curve is demonstrated. 
 These two universal curves basically determine the $dn/d\eta$.
 The energy dependence shows up through the position of the beginning of the
 central plateau.
 
 The hypothesis of limiting fragmentation has  a long history.
 For {\em hadron-hadron} collisions, 
 this hypothesis was first put forward 
 by Benecket et.al.\cite{Benecke:sh}
 and also by Feynman\cite{Feynman:ej} and Hagedorn\cite{Hagedorn:gh}.
 This idea was further developed 
 in Refs.\cite{Chou:bj} -- \cite{Hoang:1995mk}.
 
 Feynman hypothesized that as $\sqrts\to \infty$, the 
 multiplicity spectrum 
 \be
 \lim_{\sqrts\to \infty} E_p\, {dn_{hh}\over d^3 p} 
  = 
 \lim_{\sqrts\to \infty} {dn_{hh}\over dy\,d^2 p_T} 
  = f(x_L, p_T)
 \label{eq:scaling}
 \ee
 becomes independent of $\sqrts$.  
 Here $y = (1/2)\ln((E_p+p_L)/(E_p-p_L))$  is the rapidity 
 and $x_L = 2p_L/\sqrts$ is the longitudinal momentum fraction.
 If the mass of the particles is light compared to the average $p_T$, this
 expression also equals $dn_{hh}/d\eta\,d^2p_T$ where 
 $\eta=(1/2)\ln((p+p_L)/(p-p_L))$ is the pseudo-rapidity.
%
 The universal function $f(x_L, p_T)$ then totally determines
 the height of the $dn_{hh}/d\eta$ 
 and the total multiplicity
 at high energies.  
%
 
 Note that since $f(x_L, p_T)$ itself is independent of $\sqrts$,
 the height of the plateau $\left(dn_{hh}/dy\right)_{\eta=0}$
 must also be independent of $\sqrts$. 
 This also implies that
 the total multiplicity must behave like $\ymax \sim \ln s$
 where 
 \be
 \ymax = \cosh^{-1}(\sqrt{s}/2m_N) \approx \ln (\sqrt{s}/m_N)
 \ee
 is the beam rapidity and $m_N$ is the nucleon mass.
 However, up to $\sqrts = 1800\,\GeV$ the experimental data does not show
 that $(dn/d\eta)_{\eta=0}$ is saturated.
 Also proton-proton and proton-anti-proton 
 data show that the height of the plateau grows like $\ln^2 s$
 (See compilation by PHOBOS in Refs.\cite{Back:2003xk,Back:2001ae}.).
 The the total multiplicity then must grow like $\ln^3 s$.
 This is not what one would expect from Eq.(\ref{eq:scaling}).

 {The source of this discrepancy
 is the fact the strict Feynman-Yang scaling is not perfect nor is it
 supposed to be.
 The central region (or small $x_L$) is modified by 
 radiation of soft partons and
 and the multiple rescatterings of produced particles.  QCD radiative
 corrections should also give rise to the additional scale dependence 
 in $f(x_L, p_T)$\cite{Jalilian-Marian:2002wq}.

 However, within the dynamic range where Feynman-Yang scaling approximately
 holds, what should still work is the universality of $dn/d\eta$ near 
 $\eta = \ymax$, or equivalently at large $x_L$.  We should still have
 \be
 \left. {dn_{hh}\over d\eta} \right|_{\eta=\ymax+\eta'}
 &\approx &
 \int d^2p_T\, f((p_T/m_N)\, e^{\eta'} , p_T) \equiv f_U(\eta')
 \label{eq:tail_scaling}
 \ee
 where the universal function $f_U(\eta')$ is independent of $\sqrts$
 (modulo the separating scale dependence). 
 So far this is what the experimental data seem to show in both
 hadron-hadron collisions and the heavy-ion collisions.

 Physically, the existence of the limiting fragmentation is a consequence of
 having a universal large $x$ distributions in high energy hadrons
 combined with
 the short interaction range in the rapidity space\cite{Iancu:2002vu}. 
 Therefore, learning about the limiting fragmentation is equivalent to
 learning about the universal large $x$ distribution.
 
 In the popular Venugopalan-McLerran model of gluon dynamics,
 these large $x$ partons then act as the 
 color source that generates the small $x$ partons.
 Therefore establishing the validity and also the form of the limiting
 fragmentation in heavy ion collisions can provide an important input for
 the bulk dynamics of the soft degrees of freedom. 
 }


 As far as we can determine,
 the second universal curve in the transition region
 has never been studied before.
 In the following sections, we will argue that 
 the appearance of the universal transition curve may be anticipated.
 However, further study is needed to uncover the true cause for this
 universality.

 { 
 In this context, it is quite interesting that the deuteron-gold (d+Au)
 result contains the same fragmentation and transition region
 curve as the gold-gold (Au+Au) result.  This is discussed in more detail in
 section~\ref{subsec:dA}.
 }

 The rest of this paper is organized as follows.
 In section~\ref{sec:rhic}, we analyze available RHIC data.
 A simple parametrization of $d^2n/d\eta^2$ is presented and its
 consequences explicitly calculated.  The results from 
 several theoretical models including HIJING\cite{Wang:1996yf}, 
 UrQMD\cite{Bass:1998ca,Bleicher:1999xi}
 and a saturation model\cite{Kharzeev:2000ph,Kharzeev:2001gp}
 are compared against the universal curves.
 Using the two parametrizations of $dn/d\eta$ from previous sections, we
 make a prediction for LHC in section~\ref{sec:LHC}. 
 Discussions and Conclusions are given in section~\ref{sec:concl}.
 Appendix A contains details of a calculation not shown in the main
 text.  
 In Appendix B, we discuss the validity (or the lack of) the Wood-Saxon 
 form of $dn/d\eta$ sometimes used to describe the data.

\section{Experimental Limiting Fragmentation and Transition Curves}
\label{sec:rhic}

\subsection{Analysis of RHIC Au+Au}
\label{sec:my_param}
 \begin{figure}[t]
 \begin{center}
  \epsfxsize=0.7\tw
  \epsfbox{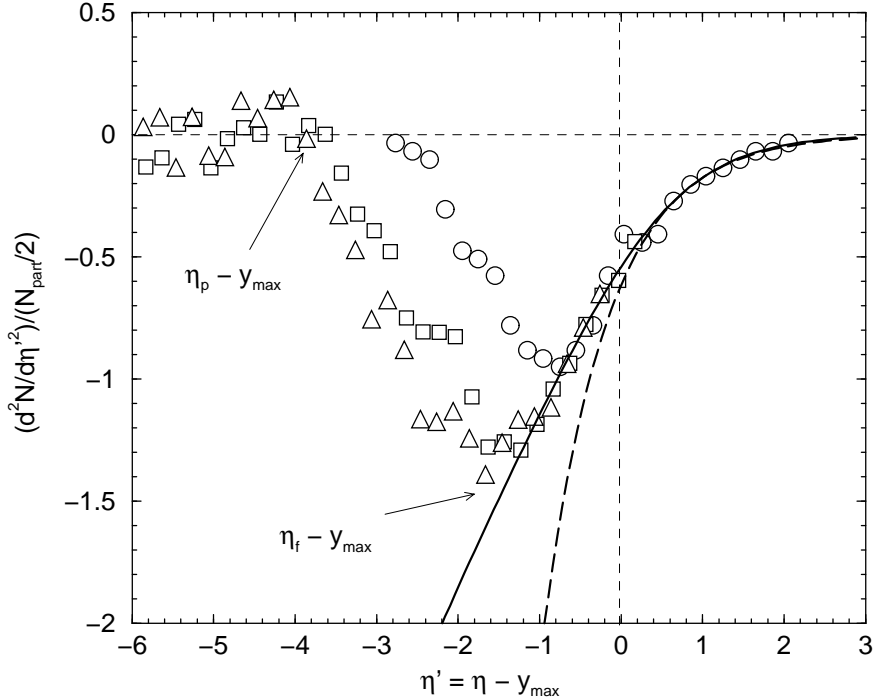}
 \end{center}
 \caption{$d^2n/d{\eta'}^2$ numerically calculated from the PHOBOS most
 central collision data\protect\cite{Back:2002wb}. 
 The triangles are for $\sqrt{s}=200\,\GeV$, the squares are for
 $\sqrt{s}=130\,\GeV$ and the circles are for 
 $\sqrt{s}=19.6\,\GeV$.  Also shown are two choices of limiting
 fragmentation functions as explained in the text.
 The two arrows mark the starting point of the plateau ($\etap-\ymax$) and the
 starting point of the fragmentation region ($\etaf-\ymax$) for
 $\sqrts=200\,\GeV$ curve.}
 \label{fig:fU}
 \end{figure}
 
 \begin{figure}[t]
 \begin{center}
  \epsfxsize=0.7\tw
  \epsfbox{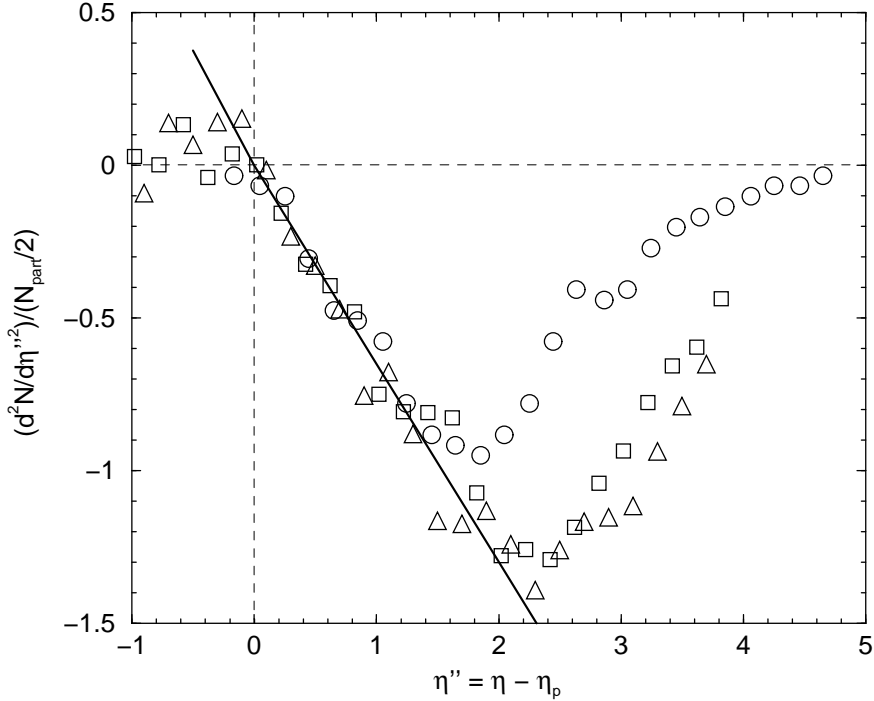}
 \end{center}
 \caption{$d^2n/d{\eta''}^2$ numerically calculated from the PHOBOS most central
 data~\cite{Back:2002wb}.  
 Here $\eta'' = \eta-\etap$
 where $\etap$ is the location of the hump in $dn/d\eta$.
 The solid line is $g_U(\eta'') = -0.65\,\eta''$. 
 }
 \label{fig:transition}
 \end{figure}
 
 \begin{figure}[t]
 \begin{center}
  \epsfxsize=0.7\tw
  \epsfbox{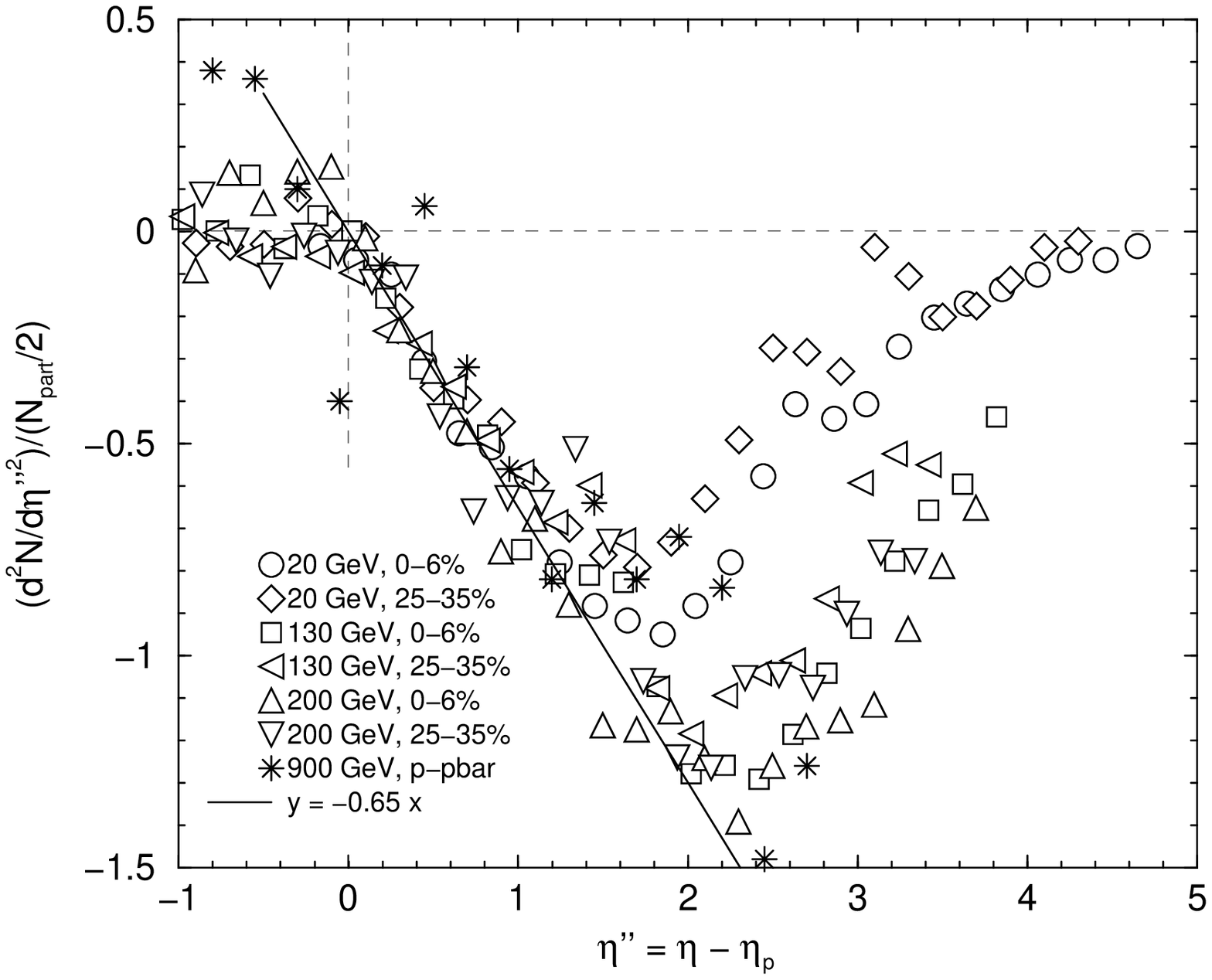}
 \end{center}
 \caption{$d^2n/d{\eta'}^2$ numerically calculated from the PHOBOS most central
 data and semi-central data~\cite{Back:2002wb}.
 Here $\eta'' = \eta-\etap$
 where $\etap$ is the location of the hump in $dn/d\eta$.
 The solid line is $g_U(\eta'') = -0.65\,\eta''$. 
 }
 \label{fig:cent_peri_shifted}
 \end{figure}

 If the universal behavior indeed extends up to 90\,\% of 
 the plateau height\cite{Back:2002wb},
 the height as well as the total multiplicity would be largely determined by
 the limiting function $f_U(\eta')$ where $\eta' \equiv \eta-\ymax$.
%
 In reality, the fragmentation region extend up to 50\,\% of
 the plateau height at RHIC energy.
 This fact is hard to see when comparing $dn/d\eta$'s 
 but becomes apparent when comparing $d^2n/d\eta^2$'s.
 In Fig.~\ref{fig:fU}, we plot $d^2n/d\eta^2$ for
 the most central collisions for 
 $\sqrt{s}= 200\,\GeV$, $\sqrt{s}=130\,\GeV$
 and $\sqrts= 19.6\,\GeV$ numerically calculated from
 the PHOBOS data\footnote{
 It is not possible to estimate experimental error bars for the slope
 without knowing the correlation between the errors.  It is likely that the
 errors in the neighboring bins are highly correlated.  
 In this paper, we assume that this is the case. 
 }.  
 One can see that there are three distinct regions (we will ignore the hump).
 The limiting fragmentation region lies to the right of the minimum of
 $d^2n/d\eta^2$ 
 ($\eta > \etaf$) in which all data points merge together.  
 To its left comes the transition region between the
 fragmentation and the plateau $(\etap<\eta< \etaf)$.  
 The zero of $d^2n/d\eta^2$
 is where the plateau begins ($\eta=\etap$).
 This is also the location of the hump in $dn/d\eta$.
 
 It is clear from this figure that the true limiting fragmentation
 region starts from about half way between the plateau and $\ymax$.
 The area of the triangular shape is the height of the plateau.
 Therefore at these energies
 the limiting fragmentation region extends up to about
 50\,\% of the maximum height.  Apparent matching of data points below
 $\etaf$ seen in $dn/d\eta'$ is due to the slow change in the slope but it is
 not a true universal behavior.   

\begin{figure}[t]
 \begin{center}
  \epsfxsize=0.7\tw
  \epsfbox{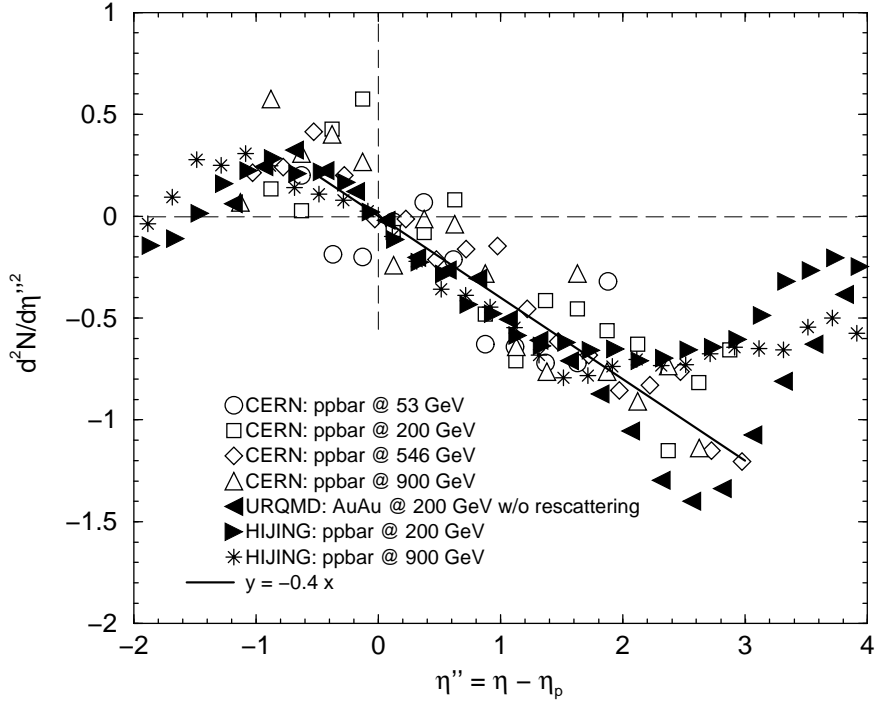}
 \end{center}
 \caption{$d^2n/d{\eta'}^2$ numerically calculated from CERN 
 data as compiled by Particle Data Group\cite{Hagiwara:fs}.
 Here $\eta'' = \eta-\etap$
 where $\etap$ is the location of the hump in $dn/d\eta$.
 The solid line is $g_U(\eta'') = -0.40\,\eta''$. 
 }
 \label{fig:ppbar_shifted}
 \end{figure}

 What is even more interesting is that the transition region also
 exhibits a universal behavior.  This is easily seen if
 one matches the zeros of $d^2n/d\eta^2$ curves (locations of the hump in
 $dn/d\eta$) as shown in Fig.~\ref{fig:transition}. 
 One can see that all data points again merge together.
 We will denote this `universal curve' as
 \be
 g_U(\eta'') \equiv \left.{d^2n\over d\eta^2}\right|_{\eta = \etap + \eta''}
 \ \ \ \ \ \ 
 (\etap < \eta < \etaf)
 \ee
 In Fig.~\ref{fig:cent_peri_shifted}, we also show the semi-central data
 from PHOBOS together with the central collision data. 
 The quality of the data is not as good as the central
 collision data, but the universal behavior is still evident. 
 We do not plot very peripheral data in Fig.~\ref{fig:cent_peri_shifted} 
 since the participant scaling seems not to have been 
 well established for them \cite{Back:2002wb}.  Instead, in
 Fig.~\ref{fig:ppbar_shifted}, we plot the result of $p\bar p$ collisions at
 various high energies as measured at CERN together with a UrQMD
 calculation from Ref.~\cite{Bleicher:1999pu} and HIJING results.
 The quality of data for these measurements are not as clean as RHIC data
 from PHOBOS.  However, there is a strong indication that there is a common
 transition curve.  There is also an indication that the slope in $p\bar p$
 $(-0.4)$ is different from the heavy ion result $(-0.65)$.
 As argued in Section~\ref{subsec:dA}, this is most likely due to the
 nuclear modification which is also supported by the HIJING and UrQMD
 results shown in Fig.~\ref{fig:ppbar_shifted}.
 From now on, we will focus our attention on the central heavy ion collisions.

%
%
 
 The shape of $dn/d\eta$ is determined by
 the functional forms of $f'_U(\eta') = df_U/d\eta'$
 and $g_U(\eta'')$ and the condition that these two curves
 meet at the transition point $\eta = \etaf$:
 \be
 g_U(\etaf-\etap) = f_U'(-\ymax+\etaf)
 \label{eq:limiting_cond_gen}
 \ee
 This is the condition that connects the behavior of the fragmentation
 region to the plateau region.
 Once the value of $\etap$ is determined by the
 zero of $d^2n/d\eta^2$, the pseudo-rapidity distribution $dn/d\eta$ is
 fully determined by $f_U$, $g_U$ and the condition 
 (\ref{eq:limiting_cond_gen}).

 \begin{figure}[t]
 \begin{center}
  \epsfxsize=0.6\tw
  \epsfbox{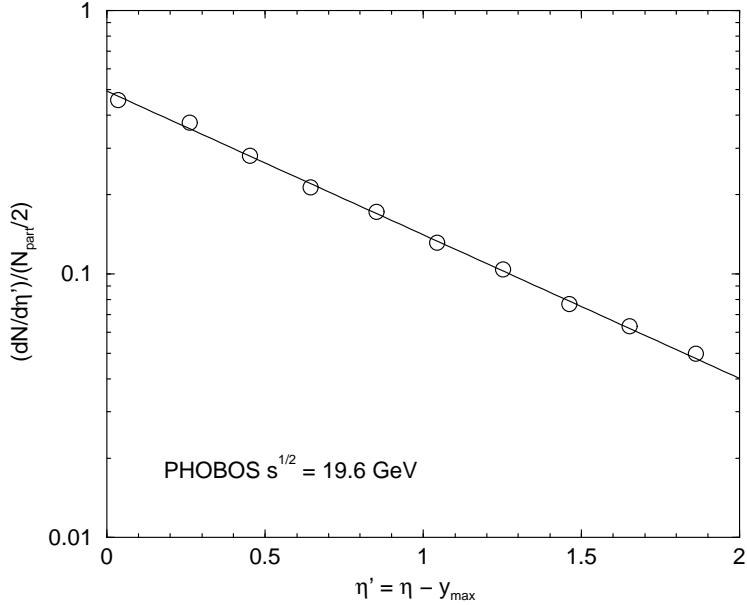}
 \end{center}
 \caption{The tail part of the
 PHOBOS data on $dn/d\eta'$ for 
             the central 6\,\% of Au+Au collisions at $\sqrt{s}= 19.6 $\,GeV.
 The straight line is $(dn/d\eta') = 0.492\, e^{-1.253\, \eta'}$. }
 \label{fig:tailfit}
 \end{figure}

 \begin{figure}[t]
 \begin{center}
  \epsfxsize=0.7\tw
  \epsfbox{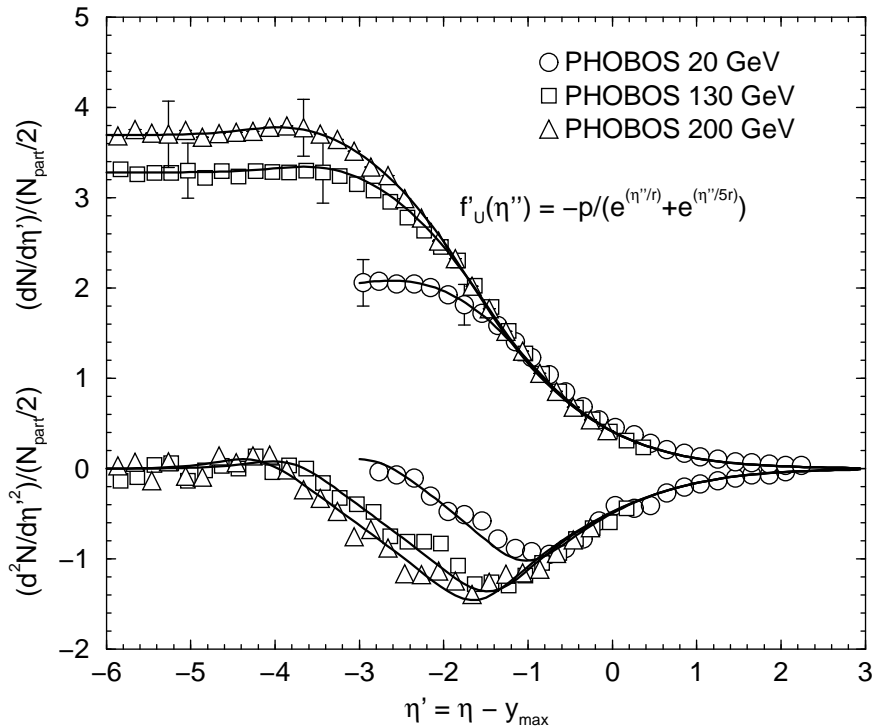}
 \end{center}
 \caption{Pseudo-rapidity distribution for 6\,\% most
 central Au-Au collisions at
 $\sqrts = 200\,\GeV$.  Data are from Ref.~\cite{Back:2002wb}.
 Lower curves are $d^2n/d\eta^2$ calculated numerically from $dn/d\eta$ data.
 The solid lines are our fits.
 Here we used 
 $\tilde{\theta}_\rho(x-x_0)
 =
 1/(1 + e^{-(x-x_0)/\rho})$ with $\rho = 0.25$
 and the parameters are set to $p = 0.95$, $1/r = 0.308$, $q = r/5$, 
 $\delta=0.3$ and $K = 0.65$.  The value of $p$ is different from 1.08
 quoted in the text because a finite value of $\rho$ compensates it a little.
 Here $\etap$ is not free but fixed by the position of the hump.
 They are at $\eta' = -3.96$, $\eta' = -3.65$ and $\eta'=-2.6$ for
 $\sqrts=200, 130, 19.6\,\GeV$ respectively.
 }
 \label{fig:new_dndeta}
 \end{figure}
 
 \begin{figure}[t]
 \begin{center}
  \epsfxsize=0.7\tw
  \epsfbox{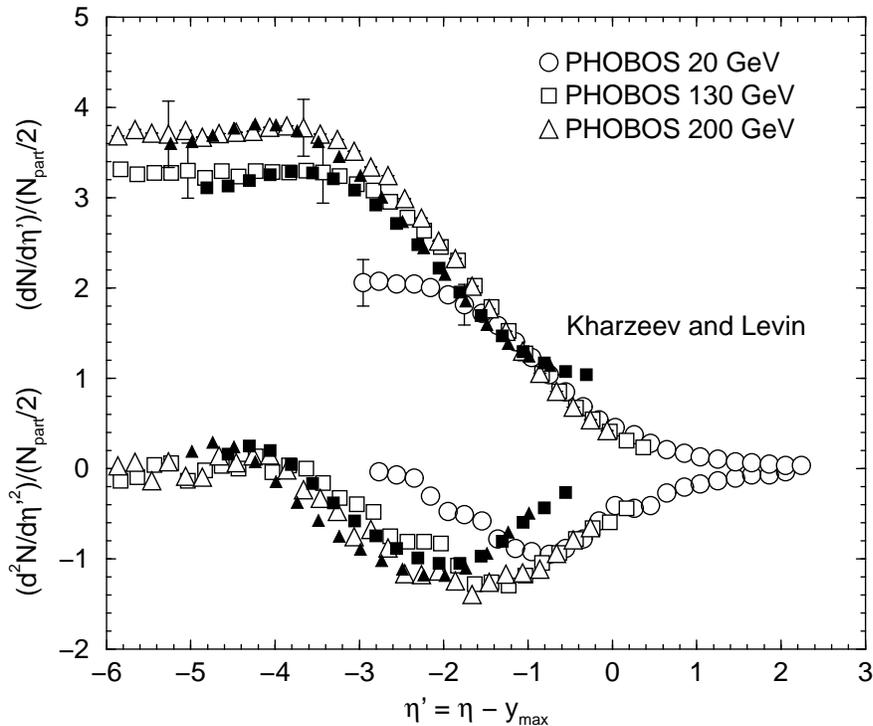}
 \end{center}
 \caption{$dn/d\eta$ and $d^2n/d\eta^2$ calculated using Kharzeev and
 Levin's result\cite{Kharzeev:2001gp}.  Open symbols are PHOBOS results and
 full symbols are the calculations.}
 \label{fig:KL}
 \end{figure}
 
 \begin{figure}[t]
 \begin{center}
  \epsfxsize=0.7\tw
  \epsfbox{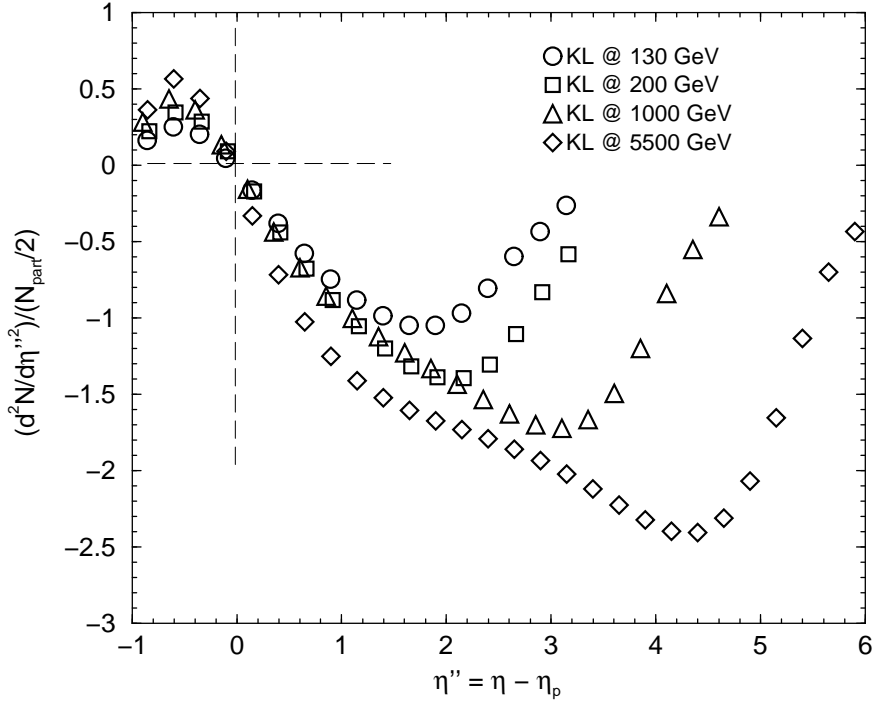}
 \end{center}
 \caption{$d^2n/d\eta^2$ calculated using Kharzeev and
 Levin's result\cite{Kharzeev:2001gp} up to the LHC energy.} 
 \label{fig:KL_d2ndy2}
 \end{figure}
 
 \begin{figure}[t]
 \begin{center}
  \epsfxsize=0.7\tw
  \epsfbox{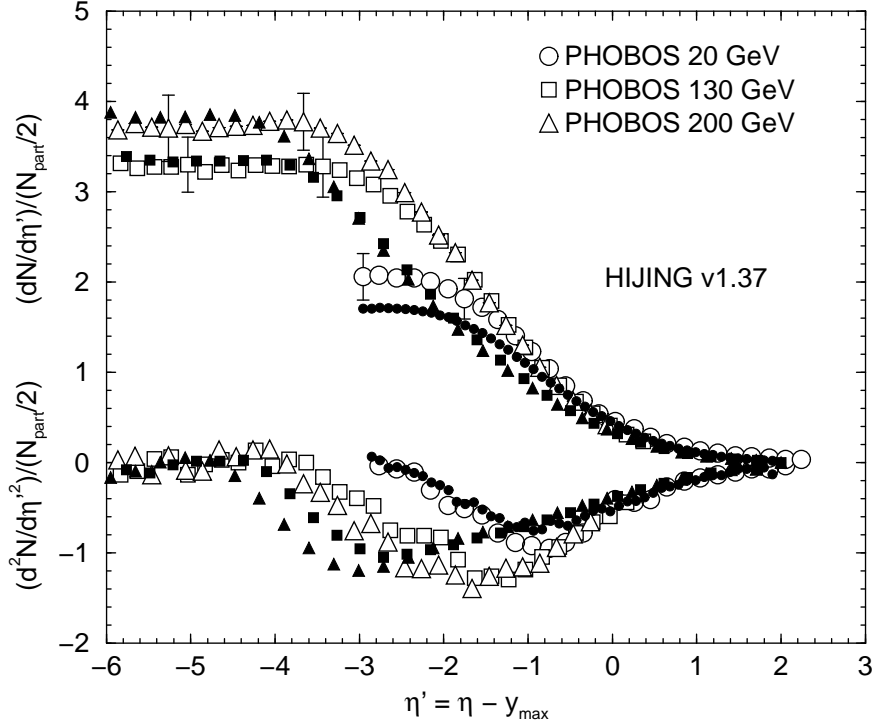}
 \end{center}
 \caption{$dn/d\eta'$ and
 $d^2n/d{\eta'}^2$ from the central PHOBOS
 data compared with HIJING v1.37 (filled symbols)
 calculations with $dE/dx = -2$\,GeV/fm
 \cite{ToporPop:2002gf}.
 } 
 \label{fig:hijing137}
 \end{figure}
 
 \begin{figure}[t]
 \begin{center}
  \epsfxsize=0.7\tw
  \epsfbox{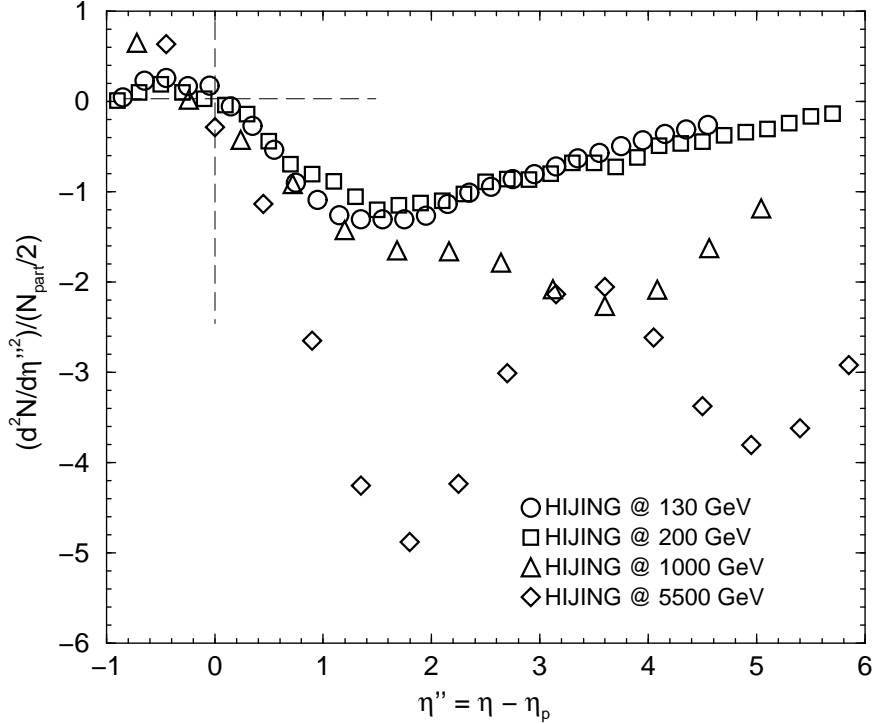}
 \end{center}
 \caption{$d^2n/d{\eta'}^2$ numerically calculated from the central PHOBOS
 data compared with HIJING v1.37 calculations.} 
 \label{fig:HIJING_d2ndy2}
 \end{figure}
 
 \begin{figure}[t]
 \begin{center}
  \epsfxsize=0.7\tw
  \epsfbox{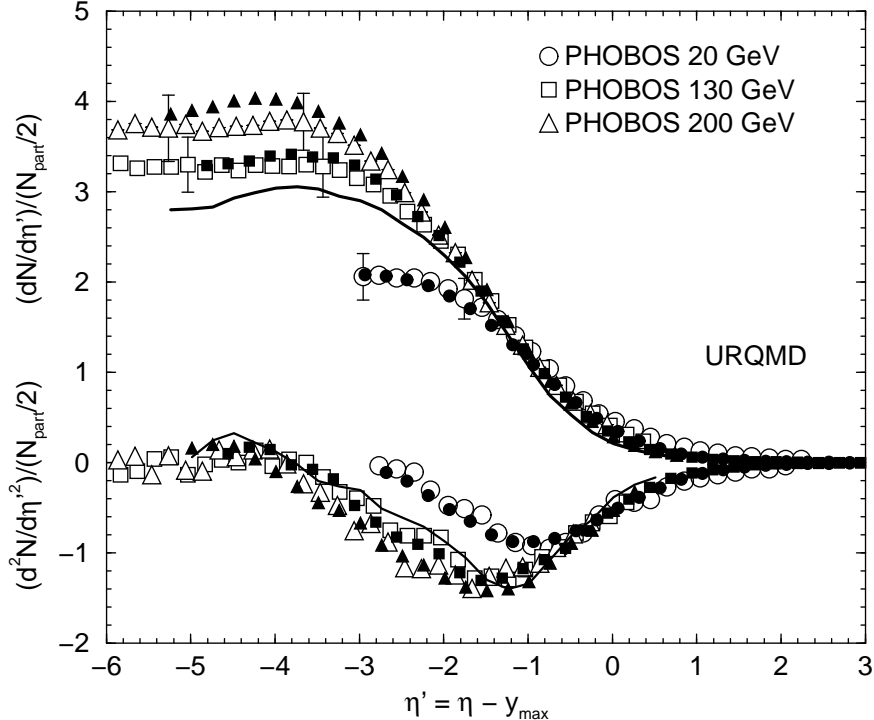}
 \end{center}
 \caption{$d^2n/d{\eta'}^2$ numerically calculated from the central PHOBOS
 data compared with UrQMD calculations (filled symbols).
 The solid line is UrQMD result for
 $\sqrt{s}=200\,\GeV$ without rescatterings.
 } 
 \label{fig:urqmd}
 \end{figure}

 \begin{figure}[t]
 \begin{center}
  \epsfxsize=0.7\tw
  \epsfbox{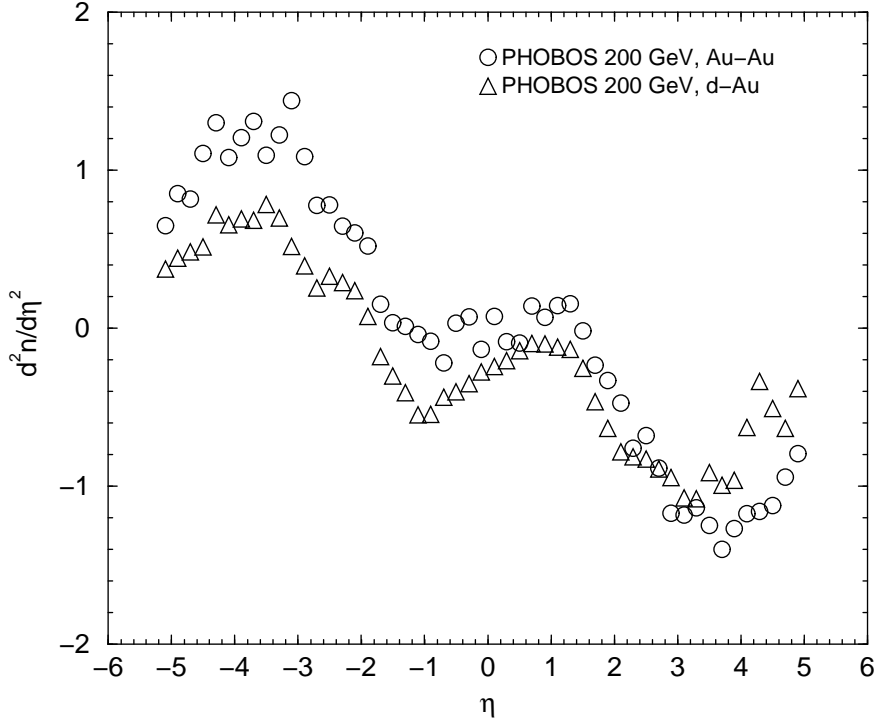}
 \end{center}
 \caption{The derivative $(d^2N/d\eta^2)/(N_{\rm part}/2)$
 with respect to $\eta$. Data are taken from the PHOBOS
 website\cite{phobosweb}.}
 \label{fig:phobos_da_d2ndeta2_part}
 \end{figure}

 \begin{figure}[t]
 \begin{center}
  \epsfxsize=0.7\tw
  \epsfbox{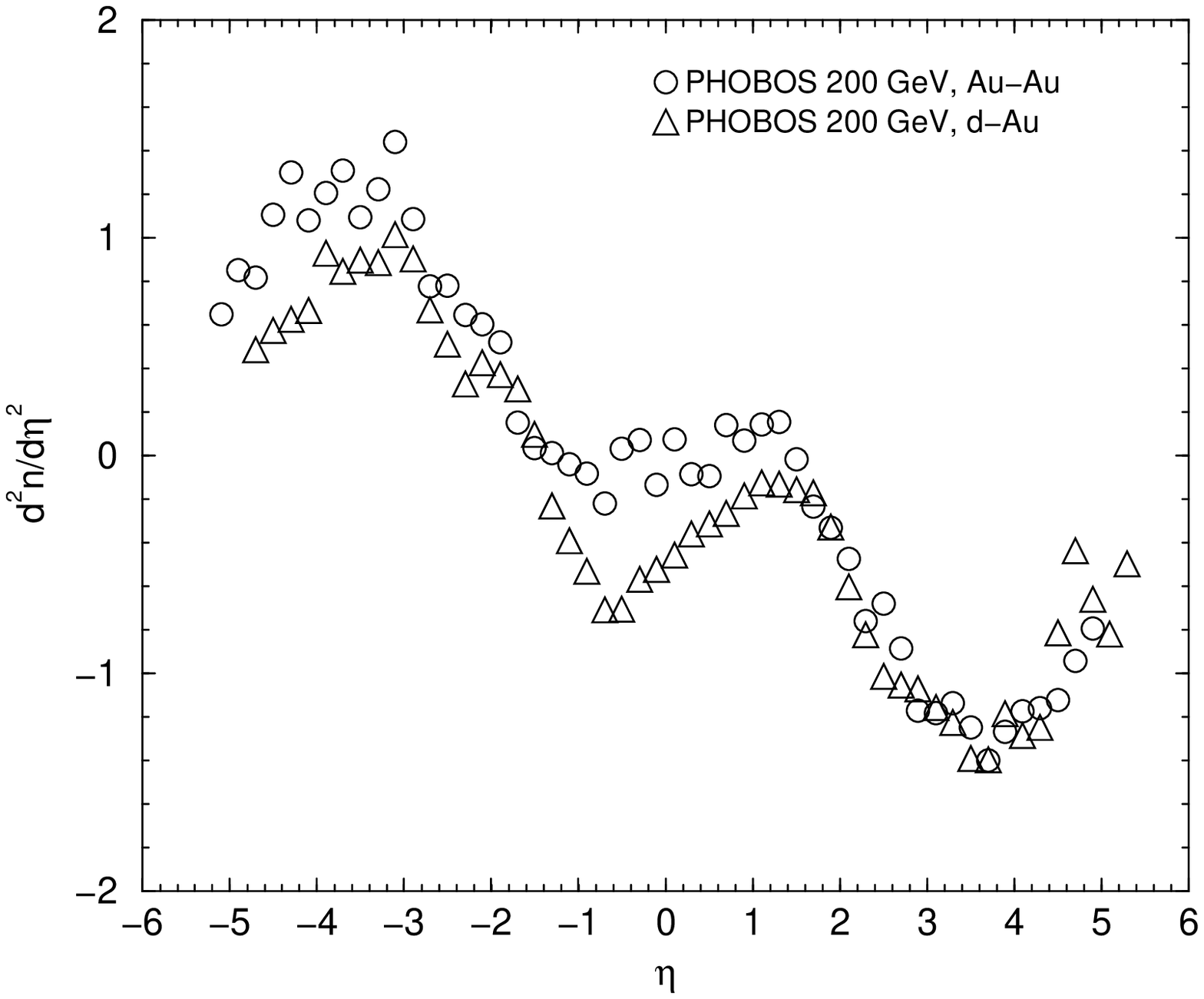}
 \end{center}
 \caption{
 The same as Fig.\protect\ref{fig:phobos_da_d2ndeta2_part} but d+Au data is
 vertically scaled by a factor of 1.3 and shifted to the right by 0.4 unit
 of pseudorapidity (2 experimental bins).
 }
 \label{fig:phobos_da_d2ndeta2}
 \end{figure}

 \begin{figure}[t]
 \begin{center}
  \epsfxsize=0.7\tw
  \epsfbox{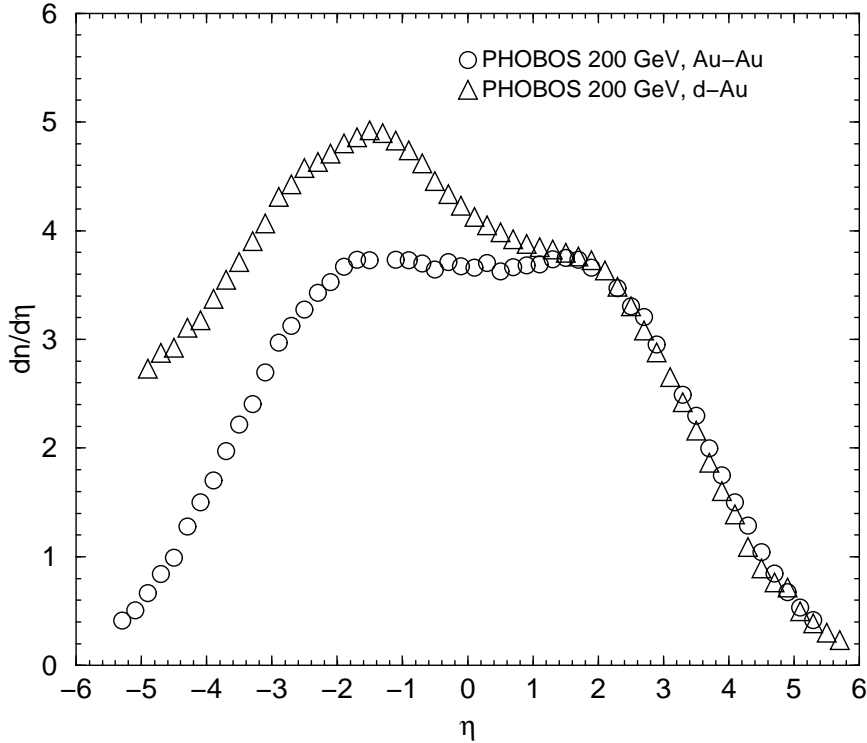}
 \end{center}
 \caption{
 The same as Fig.\protect\ref{fig:phobos_da_d2ndeta2} but for $dN/d\eta$.}
 \label{fig:phobos_da_dndeta}
 \end{figure}
 
 \begin{figure}[t]
 \begin{center}
  \epsfxsize=0.7\tw
  \epsfbox{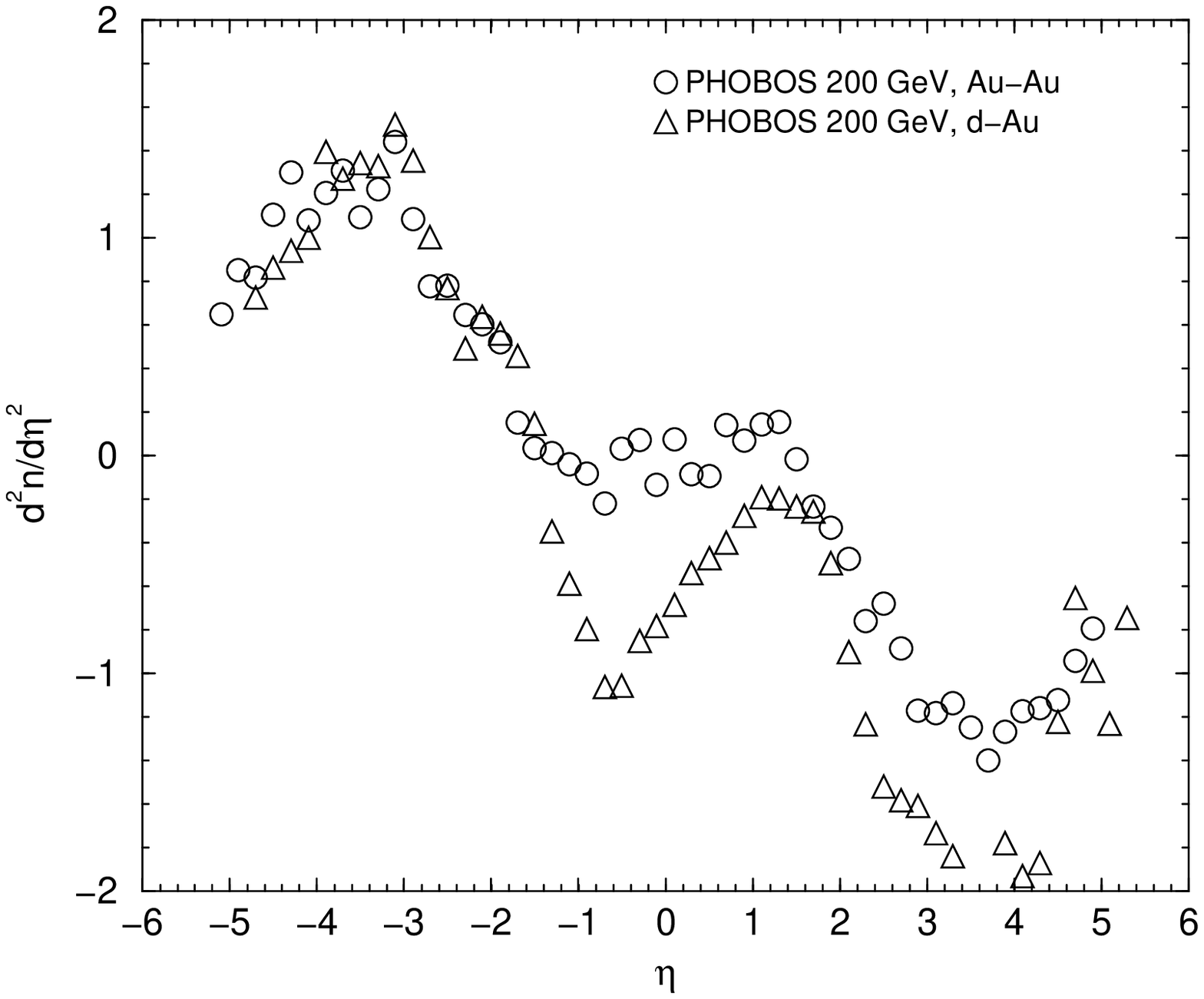}
 \end{center}
 \caption{
 The same as Fig.\protect\ref{fig:phobos_da_d2ndeta2_part} but d+Au data is
 vertically scaled by a factor of 1.9 and shifted to the right by 0.2 unit
 of pseudorapidity (1 experimental bin).
 }
 \label{fig:phobos_da_d2ndeta2_left}
 \end{figure}

 \begin{figure}[t]
 \begin{center}
  \epsfxsize=0.7\tw
  \epsfbox{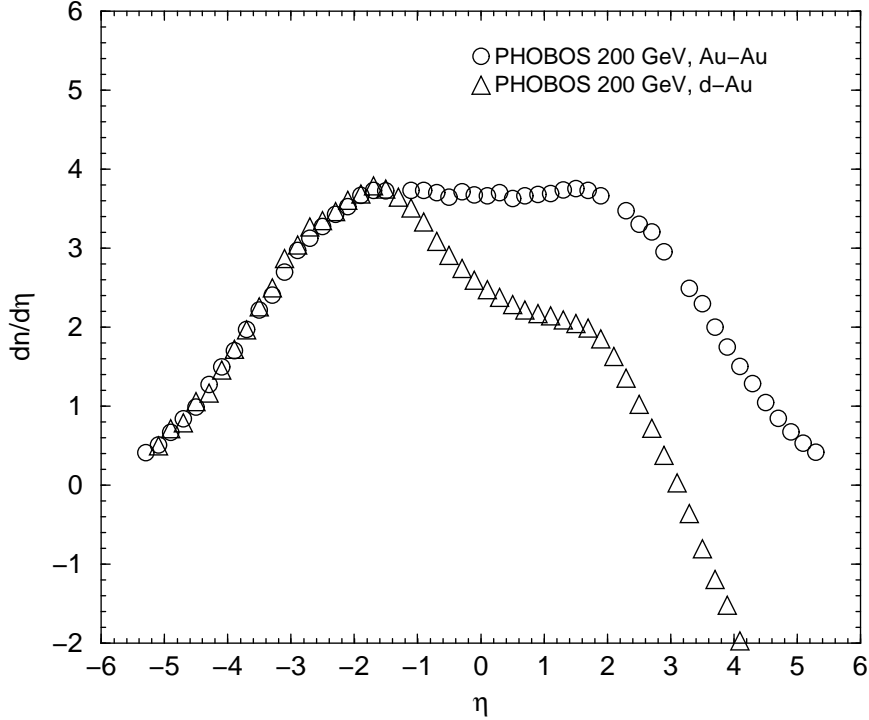}
 \end{center}
 \caption{
 The same as Fig.\protect\ref{fig:phobos_da_d2ndeta2_left} but for $dN/d\eta$.
 d+Au result is shifted vertically down by 3.6.
 }
 \label{fig:phobos_da_dndeta_left}
 \end{figure}
 
 \begin{figure}[t]
 \begin{center}
  \epsfxsize=0.7\tw
  \epsfbox{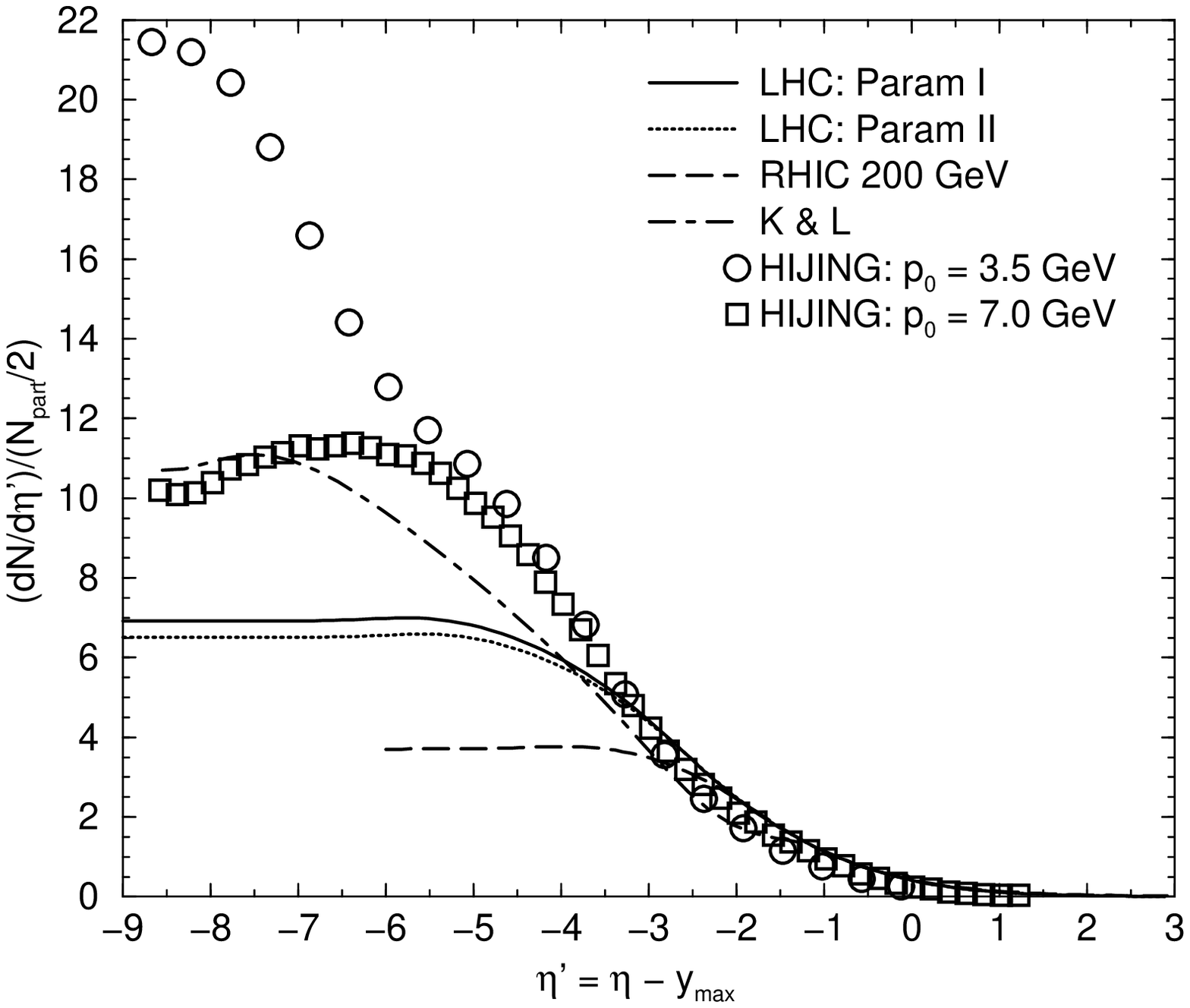}
 \end{center}
 \caption{Predictions for central collisions at LHC.  
 Also shown are parametrized RHIC 200 result.
 We set $\sqrt{s} = 5.5\,\hbox{TeV}$ and Pb+Pb for LHC.} 
 \label{fig:lhc_dndy}
 \end{figure}

 A question then arises: What are the functional forms
 of the limiting fragmentation curve $f_U$ and the transition curve $g_U$? 
 For the transition curve $g_U$, the current RHIC data 
 shown in Fig.~\ref{fig:transition}
 suggest that it is a linear function of $\eta$ with a $\sqrts$ independent
 slope.   In this paper, we take this to be true and write
 \be
 g_U(\eta'') = -K \eta''  
 \label{eq:gU}
 \ee
 where $\eta'' = \eta - \etap$.
 The value of $K$ we use is set to $0.65$ which is the slope of the
 straight line shown in Fig.~(\ref{fig:transition}).

 As for the limiting fragmentation function, 
 there is little doubt that $f_U$ is exponential for $\eta >\ymax$
 as can be clearly  seen in Fig.~\ref{fig:tailfit}. 
 But what about below $\ymax$?  
 A current theoretical analysis\cite{Jalilian-Marian:2002wq}
 relates $f_U$ to the gluon distribution function at large $x$. 
 At moderate $Q^2$, the gluon distribution function has the form
 \be
 xG(x, Q^2) \sim x^{-\lambda}\, (1-x)^n
 \ee
 With $x = e^{-2-\ymax+\eta} = e^{-2+\eta'}$\cite{Jalilian-Marian:2002wq}, 
 this means that the limiting fragmentation function should behave like
 \be
 f_U(\eta') \sim e^{-\lambda\eta'} (1-e^{-2+\eta'})^n
 \label{eq:JJM}
 \ee
 Here $-2$ in the exponent is due to the mass difference between a proton
 and a pion.
 The behavior of the expression 
 (\ref{eq:JJM}) is different from the exponential behavior shown in
 Fig.~\ref{fig:tailfit} in the $\eta' > 0$ 
 $(\eta>\ymax)$ region.  However 
 in the $\eta' < 0$ $(\eta < \ymax)$ region,
 \be
 f_U(\eta') \sim e^{-\lambda\eta'} 
 \ee
 gives a reasonable description with 
 $\lambda \sim 0.25$\cite{Jalilian-Marian:2002wq}.

 Combined, the above analysis indicate that the behavior of $dn/d\eta$
 changes from one exponential form to another exponential form when $\eta$
 crosses $\ymax$ (or $\eta'$ crosses $0$).
 We may represent such behavior with
 \be
 f'_U(\eta') 
 \equiv {df_U\over d\eta'}
 = -{p\over e^{\eta'/r} + e^{\eta'/q}}
 \label{eq:frag_param1}
 \ee
 In Fig.~\ref{fig:fU}, the solid curve corresponds to this form 
 fitted to $\eta' > -1$ portion of the combined data 
 set.
%
%
 The dashed curve corresponds to the extension of the exponential from $\eta' > 0$
 region.

 If the data points
 shown in Fig.~\ref{fig:fU} follow {\em the true universal  
 curve}, then we have no choice but to conclude that 
 $f_U$ changes its behavior once $\eta=\ymax$ is crossed.
 On the other hand, we may also consider that
 RHIC energy is not high enough for the true universal curve to manifest
 and what we see in the current data is an accident.  One such example is
 presented in Appendix~\ref{app:WS}.  For reasons explained in the Appendix, 
 this accident is unlikely.  However, only further experiments can give a
 definite verdict.

 Using Eq.~(\ref{eq:gU}) for $g_U$, we write for $\eta > 0$  
 \be
 {d^2n\over d\eta'^2}
 =
 -\left( {p\over e^{\eta'/r} + e^{\eta'/q}} \right)\,
  \tilde{\theta}_{\rho}(\eta'+\chif)
 -K(\eta'+\chip)\,\tilde{\theta}_\rho(-\chif-\eta') \, 
 \tilde{\theta}_\rho(\eta'+\chip-\delta)
 \label{eq:model1}
 \ee
 where $\eta' = \eta - \ymax$ and
 $ \tilde{\theta}_\rho(x)$
 is the smeared $\theta$ function with $\lim_{\rho\to
 0}\tilde\theta_\rho(x)=\theta(x)$.
 The minimum of $d^2n/d\eta'^2$ is located at $\eta' = -\chif = \etaf -\ymax$
 and the hump of $dn/d\eta'$ is located at $\eta' = -\chip = \etap - \ymax$.
 The parameters $\delta$ and $\rho$ control the height and the width
 of the hump. The parameter $K = 0.65$ is the slope of the transition
 region.  
 By fitting the combined $\eta' > -1$ data, we get 
 $p = 1.08$, $1/r = 0.308$ and $1/q = 1.566 \approx 5/r$.
 We mention here that this
 $1/r$ value is fairly close to the value of a similar coefficient obtained
 from saturation model studies\cite{Jalilian-Marian:2002wq,Kharzeev:2000ph,
 Kharzeev:2001gp,Kharzeev:2002pc,Dumitru:2002wd,Kovchegov:1999ep}.
 The shape of $dn/d\eta$ obtained from Eq.(\ref{eq:model1}) as well as
 $d^2n/d\eta^2$ itself is shown in Fig.~\ref{fig:new_dndeta}.  In the
 figure, $p = 0.95$ is used because having finite $\rho$ changes the slope 
 a little.
 
 The transition between the fragmentation region and the transition region
 happens at $\eta' = -\chif$ where the two lines meet.  This yields the
 following condition
 \be
 K(\chip - \chif) = {p\over e^{-\chif/r} + e^{-\chif/q}}
 \approx p\, e^{\chif/r}
 \label{eq:limiting_cond1}
 \ee
 where the approximation works for large $\chif/r$.  
 This is the condition which relates the limiting fragmentation to
 the plateau and ultimately determines the size of the fragmentation region.
 For large $\chif/r$, the solution is given by
 \be
 \chif \approx \chip - r\, W(e^{\chip/r} p/Kr)
 \label{eq:chif_approx}
 \ee
 where the Lambert function solves $w = W(w)\exp(W(w))$.
%
 With the values of the
 parameters from above, the approximation 
 (\ref{eq:chif_approx}) is good within 1\,\% for RHIC and
 LHC but not for SPS.
 For future reference, we note that
 for large $w$, $W(w) \approx \ln w - \ln\ln w$.
 Hence 
 \be
 \chif 
  \sim \ln \chip \ll \chip
  \label{eq:chif_ll_chip}
 \ee

 Integrating Eq.(\ref{eq:model1})
 from $\infty$ to $\eta'$ gives the rapidity distribution $dn/d\eta'$.
 Numerical integration yields excellent description of the existing data
 as shown in Fig.~\ref{fig:new_dndeta}.
 Unfortunately,
 the form of $f_U'(\eta')$ in Eq.(\ref{eq:frag_param1}) does not allow 
 analytic integration in general.  
 However, note that $1/q \approx 5/r$.
 If $1/q = 5/r$, the necessary integration can be carried out in  
 the sharp $\theta$-function limit ($\tilde{\theta}_\rho\to \theta$). 
 The resulting form is analytic but not very illuminating.
 Details can be found in Appendix~\ref{sec:integral}.  

 Now consider the height of the plateau, $(dn/d\eta)_0$.
 Note that the fragmentation region $d^2n/d\eta^2$
 behaves exponentially while the transition region $d^2n/d\eta^2$ behaves
 linearly in $\eta$.  Therefore in the large $\ymax$ limit, 
 the contribution from the transition region dominates 
 in $(dn/d\eta)_0$.
 Physically, this is what one would expect.  At high enough energies, the
 dynamics of the central plateau region and the dynamics of the limiting
 fragmentation region should decouple and 
 the height of the central plateau should not depend
 much on the exact form of $f_U$.
 The height of the plateau in the large $\ymax$ limit is then given by
 \be
 \left(dn\over d\eta\right)_0
 \approx
{K\over 2}\left(\chif-\chip\right)^2 + O(\ymax)
 \label{eq:dndetaz_exp}
 \ee
 This implies
 \be
 \left(dn\over d\eta\right)_0 < {K\over 2} \ln^2(\sqrt{s}/m_N) \ee
 since $\etap < \etaf < \ymax$ and $\chif \sim \ln\chip$.
 Integrating once more, the total multiplicity can be obtained as
 \be
n_{\rm total}
& \approx &
{K\over 3}(\etaf - \etap)^2(2\etaf + \etap)
 + O(\Ymax^2)
\label{eq:ntot_exp}
\ee
which implies
\be
n_{\rm total} < {2K\over 3} \ln^3 (\sqrts/m_N)
\label{eq:ntot_bound}
\ee

From Eqs.(\ref{eq:dndetaz_exp}) and (\ref{eq:ntot_exp})
we conclude that the central plateau {\em cannot} rise
faster than $y_{\rm max}^2$ or $\ln^2 s$ and the total multiplicity {\em cannot}
rise faster than $y_{\rm max}^3$ or $\ln^3 s$.  The only possible way to get
faster dependence is to have $\sqrts$ dependent $K$, or faster rising $g_U$
(for instance an exponential).
Judging from Fig.~\ref{fig:ppbar_shifted}, this is not
likely up to 900\,GeV.
Also there is an additional evidence from the CDF 
collaboration~\cite{Abe:1989td} that up to $\sqrt{s} = 1.8\,\hbox{TeV}$, the
central plateau in $p\bar p$ collisions rises only as fast as $\ln^2 s$.

In many current models of heavy ion collisions,
$n_{\rm total}$ grows faster than $\ln^3 s$. 
For instance, Ref.\cite{Kharzeev:2001gp} 
has $n_{\rm total} \sim s^{\lambda/2}$ 
and the $e^+ e^-$ model\cite{Back:2003xk}
has $n_{\rm total} \sim e^{c\sqrt{\ln s}}$ where $\lambda$ and $c$ are
constants.  Parametrization of $pp$, $pA$
and $AA$ data up to the SPS energy by Gazdzicki and 
Hansen\cite{Gazdzicki:ih} gives $N_{\rm ch} \propto s^{1/4}$.
At present energies, these are indistinguishable from polynomials in
$\ln s$.  However, as will be presented shortly, LHC will be able to
tell whether the bound (\ref{eq:ntot_bound}) indeed holds for high energy
heavy ion collisions.

 At this point, we can attempt a partial explanation of
 the appearance of the universal transition curve.  
 Suppose that 
 as the collision energy becomes larger
 the dynamics of plateau region largely decouples from the dynamics of
 the fragmentation region. 
 This is certainly the case for the $f_U$ and $g_U$ given in this section as
 indicated by Eq.(\ref{eq:chif_ll_chip}).  
 Eq.(\ref{eq:chif_ll_chip}) implies that in the large $\ymax$ limit,
 $\etaf \gg \etap$ and
 \be
 \etaf = \ymax + O(\ln \ymax)\, .
 \ee
 This is a consequence of having
 an exponential fragmentation curve and a polynomial transition curve.
 Note that the functional form of $g_U$ and $f_U$ enters only through 
 the logarithmic correction.

 At RHIC energies, the area under $f_U$ and $g_U$ looks like an isosceles
 triangle.
 This is because $\ln \ymax$ is still not that small compared to $\ymax$.
 However since an exponential rises fast, the area will look more
 and more like a right triangle as the energy grows and the area will become 
 dominated by the transition part:  
 \be
 \left(dn\over d\eta \right)_0
 & \approx &
 \int^{\etap}_{\etaf} d\eta\, g_U(\eta-\etap)
 \non
 & = &
 -\int^{\etaf-\etap}_{0} d\eta''\, g_U(\eta'')
 \approx
 -\int^{\ymax}_{0} d\eta''\, g_U(\eta'')\, .
 \label{eq:transition_dom}
 \ee
 Therefore, to leading order in $\ymax$,
 $\left(dn/d\eta \right)_0$ is a function of $\ymax \approx
 \ln(\sqrt{s}/m_N)$ {\em and} it is independent of the functional form of
 the fragmentation curve $f_U$.  
 Denoting the functional dependence as
 $(dn/d\eta)_0 = S(\ymax)$,
 the universality of $g_U$ follows if  
 the following relationship holds
 \be
 g_U(\eta'') \approx -{dS(\eta'')\over d\eta''}\, .
 \label{eq:gUdS}
 \ee
 Once the dependence of $\left(dn/d\eta\right)_0$
 on $\sqrts$ is given, $g_U$ is totally determined and it is indeed universal up to logarithmic corrections.
 The relation (\ref{eq:gUdS}) certainly holds for
 Eqs.(\ref{eq:gU}) and (\ref{eq:dndetaz_exp}) when $\ymax \gg 1$.
 
 The hole in this argument is that  
 the relationship (\ref{eq:transition_dom}) does not automatically imply
 Eq.(\ref{eq:gUdS}).  For instance, 
 suppose $S(\ymax) = a y_{\rm max}^2$.  In this case, any
 \be
 g_U(\eta') = {a\,n \,y_{\rm max}^{2-n}}\, (\eta'')^{n-1}
 \ee
 with $n > 1$ satisfies Eq.(\ref{eq:transition_dom}).
 Unless $n = 2$, however, $g_U(\eta'')$ depends
 on $\ymax$ and hence it is not universal. 
 Surprising fact is that the data seems to suggest $n$ is indeed 2 or at
 least very close to it.

 The relationship (\ref{eq:gUdS}) is remarkable.  It relates an observable
 that is a function of colliding energy 
 to an observable that is a function of the pseudo-rapidity at any 
 fixed energy.  Unfortunately, energies probed so far are too small
 for this to manifest.  As seen in Figures
 \ref{fig:transition}-\ref{fig:ppbar_shifted}, the transition region is not 
 truly dominant yet. 
 However, we should be able to test this relationship at LHC. 

 It is also instructive to compare some theory curves with RHIC data
 as shown in Figs.\,\ref{fig:KL}-\ref{fig:urqmd}.
 As shown in Figs.\ref{fig:KL} 
 the saturation model by Kharzeev
 and Levin\cite{Kharzeev:2001gp} gives a good description of the 
 plateau and the transition region at RHIC energy although the fragmentation
 region is badly off.  However, since the model is based on small $x$
 picture, it is not supposed to be valid in the fragmentation region.
 From the expression of $dn/dy$ given in Ref.\cite{Kharzeev:2001gp},
 it is clear that the transition curve obtained by Kharzeev and Levin
 is exponential and
 this form of $dn/dy$ does not satisfy the relation (\ref{eq:gUdS}).
 The reason Fig.~\ref{fig:KL_d2ndy2} shows approximate universal behavior up
 to $\sqrt{s} = 1000\,\GeV$ is $\lambda$ is small.
 At LHC energy, the violation of the universality is clearly seen for this
 model.

 HIJING results \cite{ToporPop:2002gf,Wang:2000bf}
 with shadowing and a parton energy loss of
 $dE/dx = -2$\,GeV/fm and an energy dependent scale
 parameter $p_0$ as considered in \cite{Li:2001xa} 
 are shown in Fig.\ref{fig:hijing137}.
 It is quite evident in the $d^2n/d\eta^2$ plot that the fragmentation
 region dominates in HIJING.  
 Again to test the transition curve universality, one must go beyond the
 RHIC energy.  Fig.\ref{fig:HIJING_d2ndy2} shows $d^2n/d\eta^2$ up to the LHC
 energy.  From the figure, it is quite clear that HIJING does not contain 
 a universal transition curve.  Furthermore, at higher energies,
 the central region develops a bump instead of a plateau.
 This feature is due to the abundance of the minijets.  
 As can be seen in
 Fig.~\ref{fig:lhc_dndy}, by enhancing the parameter $p_0$ (equivalently,
 reducing the number of minijets) HIJING becomes closer to the other models.  
 But the transition region universality is clearly not a feature in the
 HIJING model.
 
 On the other hand, it is quite striking that the default UrQMD results get
 both $dn/d\eta$ and $d^2n/d\eta^2$ right at RHIC energies.
 It is also significant that without rescatterings, UrQMD does not  
 describe the data well.   Why UrQMD results in universal transition curve
 is not yet clear. 

\subsection{Analysis of RHIC d+Au}
\label{subsec:dA}
 Recently the PHOBOS collaboration published the result of measuring the
 pseudorapidity distribution of produced particles in the deuteron-gold
 (d+Au) collisions at $\sqrt{s} = 200\,\GeV$\cite{Back:2003hx}.
 At a first glance, it would seem that there is no common feature at all
 between the d+Au $dn/d\eta$ and Au+Au $dn/d\eta$, especially if one just
 looks at the participant-scaled results.
 However, when dealing with very asymmetric systems such as d+Au, one must be
 careful about the scaling behavior.  
 As can be easily shown in a simple wounded nucleon model, 
 the scaling of produced particles in the heavy ion side and the d side
 should be different.  
 The number of wounded nucleons in the heavy ion side depends on the linear
 size of the heavy ion whereas the d side always have 1 or 2 wounded
 nucleons.
 Hence, the multiplicity in the heavy ion side should have an additional
 scale factor $\sim A^{1/3}$ compared to the d side.

 To see whether there is a common feature between Au+Au and d+Au
 or not, again it is much better to look
 at the derivative $(d^2N/d\eta^2)/(N_{\rm part}/2)$ as shown in 
 in Fig.\ref{fig:phobos_da_d2ndeta2_part}.
 Judging from this figure, 
 it is clear that there {\em is} a common feature.
 To bring it out more clearly, we vertically scale the d+Au result by a
 factor of 1.3 and shift it horizontally by 0.4 unit of rapidity (or 2
 experimental bins).
 This results in Figs.~\ref{fig:phobos_da_d2ndeta2} and
 \ref{fig:phobos_da_dndeta} which leaves no room
 for doubt that the shape of $dN/d\eta$ for $\eta> 1.5$ is common to both
 Au+Au and d+Au results.
 It is also interesting to see that different scaling 
 (additional
 factor of 1.5 compared to 
 Fig.\ref{fig:phobos_da_d2ndeta2} and the rapidity
 shift of $0.2$ (or 1 experimental bin) instead of 0.4) 
 brings the Au side of
 the spectrum together as shown in Fig.\ref{fig:phobos_da_d2ndeta2_left}.
 and Fig.\ref{fig:phobos_da_dndeta_left}.
 Again, there is no room for doubt that there is a common curve.
 This implies that beside a constant component, the shape of $dn/d\eta$ for
 both Au+Au and d+Au is simply related by scaling even in the Au side.

\section{Prediction For LHC}
\label{sec:LHC}

Given the two forms of parametrization considered in the previous sections,
it is possible to extrapolate and predict what should happen at LHC where
$\sqrt{s} = 5500\,\GeV$ and $\ymax = 8.68$.
To do so, we need to parametrize the functional form of $\etap$ or $\etaf$.
The value of $\etap$ in the model from section~\ref{sec:my_param} is not
a free parameter.  The position of the
hump clearly visible in $d^2n/d\eta^2$ as a zero is the value of $\etap$.
From the PHOBOS data, one gets, $\ymax-\etap = 3.96, 3.65$ for
$\sqrts=200, 130\,\GeV$.  For $\sqrts = 19.6\,\GeV$ case, the data has
$\ymax-\etap=2.75$.  However, with our model, $\ymax-\etap=2.6$ describes
the data better. 

By fitting the above values of $\etap$ with 
$\ymax - \etap = \lambda y^\nu + \beta\ln y$ and
$\ymax - \etap = \lambda y^\nu + C$, 
we get\footnote{
Since we have only three data points, one cannot fit the full 
$\ymax - \etap = \lambda y^\nu + \beta\ln y + C$.}
\be
\ymax - \etap = \left\{
\begin{array}{ll}
0.60 + 0.73\, \Ymax^{0.91} 
& \ \  (\hbox{Model I})
\\
0.33 \ln \ymax + 0.96\, \Ymax^{0.75}
& \ \ (\hbox{Model II})
\end{array}
\right.
\label{eq:params}
\ee
These two parametrizations do not differ much up to $\ymax = 10$. 
The LHC predictions from these two parametrizations are given in
Table~\ref{table:models}\footnote{
To predict the height more accurately, we need to remember that 
Eq.(\ref{eq:dndetaz_exp}) neglects some part of the tail contribution.}. 
The shape of $dn/d\eta$ is given in the Fig~\ref{fig:lhc_dndy} together with
the Kharzeev-Levin prediction and the HIJING predictions with two different
minimum minijet energies.
The results obtained for $p_0=3.5\,\GeV$ suggested in reference
\cite{Li:2001xa}, is clearly very different from other models.
Increasing the mini-jet scale parameter
to a higher value  $p_0=7.0\,\GeV$ brings it to a better agreement with
other models.
Only data from LHC will allow us to draw a definite conclusion and to choose 
the right value for this parameter $p_0$.

One striking feature is that in both the Kharzeev-Levin model and the HIJING
model with $p_0 = 3.5\,\GeV$, 
the central plateau disappear.  This is mainly due to the fact that
these models do not contain rescatterings of the secondaries and hence
cannot not undergo a Bjorken-like expansion.

\begin{table}[t]
 \begin{tabular}{|c|c|c|c|}
 \hline
 & $\ymax-\etap$ & $(dn/d\eta)_0$ & $n_{\rm total}$ \\
 \hline
 Model I  & 5.8 & 6.9 & 87\\
 \hline
 Model II  & 5.6 & 6.5 & 83\\
 \hline
 K \& L  & -- & 10.7 & 110\\
 \hline
 HIJING w/ $p_0=3.5\,\GeV$/c & -- & 21.4 & 160\\
 \hline
 HIJING w/ $p_0=7.0\,\GeV$/c & -- & 11.6 & 100\\
 \hline
 \end{tabular}
 \caption{Predictions for LHC central collisions.
 We set $\sqrt{s} = 5.5\,\hbox{TeV}$ and Pb+Pb for LHC.} 
\label{table:models}.
\end{table}

\section{Discussions and Conclusions}
\label{sec:concl}

 In this paper, we showed that there exists a second universal behavior
 in the rapidity distribution of produced particles.
 The data we have analyzed clearly
 indicate that $d^2n/d\eta^2$'s 
 in the transition region taken at different energies
 follow a common curve.
 This is not easy to see when comparing $dn/d\eta$'s but clearly seen
 when comparing $d^2n/d\eta^2$'s. 
 
 We emphasize here that any model that purports to describe the rapidity
 distribution in the whole rapidity space must be able to reproduce not only
 the limiting fragmentation curve, but also the universal transition curve.

 The existence of the two universal curves implies that
 the shape and the size of the rapidity distribution itself is mostly
 determined by (i) the limiting fragmentation curve $f_U$, 
 (ii) the universal transition curve $g_U$ and 
 (iii) the starting point of the plateau region $\etap$. 
 Non-trivial physics resides in the $\sqrts$ dependence of $\etap$ or
 equivalently the size the central plateau. 

 The physics behind the limiting fragmentation curve is well known to be
 the Feynman-Yang scaling which states that at high energy, large $x_L$
 behavior of an inclusive cross-section is independent of $\sqrts$. 
 In this paper we argued that the physics behind the 
 universal transition curve is in fact the decoupling between the dynamics
 of the plateau and the fragmentation region.
 We note that it is also intriguing that perhaps a connection to the
 gluon parton distribution can be made.  In saturation models,
 $dn/d\eta$ is related to the gluon distribution
 function\cite{Kharzeev:2000ph,Jalilian-Marian:2002wq,Kharzeev:2001gp}.
 In this case, the universal transition curve puts a severe restriction on
 the behavior of the gluon distribution at moderate $x$.

 In this work, we found that the universal transition curve is linear in
 $\eta'' = \eta-\etap$ based on 
 $\sqrts=20 - 200\,\GeV$ RHIC data and UA5 $p\bar p$ data.
 A consequence of having a linear transition curve is that the plateau
 height $(dn/d\eta)_0$ cannot grow faster than $\ln^2 s$ and the total
 charged multiplicity cannot grow faster than $\ln^3 s$.  This polynomial
 behavior in $\ln s$ is maintained if the transition curve is polynomial in
 $\eta''$.  A power law growth $(dn/d\eta)_0 \sim s^\lambda$ or an $e^+ e^-$
 type exponential growth is possible only 
 if the transition curve is exponential.  The available data
 does not show such exponential behavior.  It does not, of course, rule out
 a change in the behavior at higher energies.  As shown in the last section,
 these possibilities can be clearly distinguished at LHC. 

 What we have found in this study also impacts hydrodynamic studies.
 As can be seen in Eq.(\ref{eq:params})
 the size of the plateau does not grow fast. Moreover as
 $\ymax \to \infty$, $\etap/\ymax \to 0$.  
 Therefore, the relative region of validity for 2-D hydrodynamic calculation
 shrinks as the energy goes up and the need for 3-D hydrodynamic calculation
 becomes greater.  Furthermore, the existence of the universal transition
 curve will tightly constrain the longitudinal evolution of the hydrodynamic
 system.  

{
 We have also analyzed the deuteron-gold result from RHIC
 and found that there is a
 single common curve that determines the shape of $dN/d\eta$ 
 for both d+Au and Au+Au cases. 
 A few conclusions can be drawn from our analysis.
 First of all, the different scaling factors for the deuteron side
 and the gold side indicate 
 that the scaling of d+Au system is more complex than a simple participant
 scaling.
 This implies that using a simple participant scaling 
 can potentially mislead the comparision between 
 the d+Au result and Au+Au result. 
 This is especially significant for the Au side
 where there appears to be a constant component on top of the two universal
 curves discussed in this paper.  Again, these facts are much more
 transparent if one compares $d^2N/d\eta^2$.
 
 Second, the existence of a function common to both Au+Au and d+Au indicates
 that the dynamics of the transition region and the fragmentation region
 in the Au+Au case
 cannot depend much on the final state interactions.  
 It can only depend on initial state parton-nucleus dynamics.  
 Especially, whether
 or not a hot and dense system is formed in Au+Au collisions does not
 influence the shape of $dN/d\eta$ beyond the plateau region. 

 The exact physical meaning of the rapidity shifts and the constant
 component in the d+Au data are under investigation.
}

\acknowledgments{The authors thanks J.Jalilian-Marian, R.Venugopalan, S.Bass,
C.S.Lam, S.Das Gupta, C.Gale, U.Heinz
and J.H.Lee for suggestions and discussions. 
M.B.~thanks GSI, DFG and BMBF for support.
S.J.~and V.T.P~are supported in part by the Natural Sciences and
Engineering Research Council of Canada and by le Fonds 
Nature et Technologies of Qu\'ebec.  
S.J. also 
thanks RIKEN BNL Center and U.S. Department of Energy [DE-AC02-98CH10886] for
providing facilities essential for the completion of this work.
}

\appendix

\section{Integral over $df_U/d\eta'$}
\label{sec:integral}

 To calculate $dn/d\eta$, we need
 \be
 \int_{\infty}^{\eta'} d\eta''\,  {df_U\over d\eta''}
 & = &
 -p\int_{\infty}^{\eta'} d\eta''\,
 {1\over e^{\eta'/r} + e^{5\eta'/r}}
 \ee
 Changing variable to $y = e^{\eta''/r}$ yields
 \be
 -p\, r \int_{\infty}^{e^{\eta'/r}} {dy\over y^2(1 + y^4)}
 & = &
 -p\, r
  \int_{\infty}^{e^{\eta'/r}} dy\,
  \left(
  {1\over y^2}
  -
  {y^2\over 1+ y^4}
  \right)
  \ee
These integrals can be found in integral tables, for instance
Ref.\cite{integ:GR}.  We get
\be
\int_{\infty}^{\eta'} d\eta''\,  {df_U\over d\eta''}
& = &
p\,r\,
\Bigg\{
e^{-\eta'/r} 
+
{p\over 2\sqrt{2}}
\left(
\tan^{-1}\left(1+\sqrt{2}e^{\eta'/r}\right)
-
\tan^{-1}\left(1-\sqrt{2}e^{\eta'/r}\right)
-\pi
\right)
\non & & {}
\qquad
-
{1\over 2}\log\left(
1 + \sqrt{2} e^{-\eta'/r} + e^{-2\eta'/r}\over
1 - \sqrt{2} e^{-\eta'/r} + e^{-2\eta'/r}
\right)
\Bigg\}
\ee

 \section{Universal Fragmentation Condition for Wood-Saxon} 
 \label{app:WS}

 A popular choice of parametrization for
 $dn/d\eta$ is the Wood-Saxon function.
 Many models for hadron-hadron collisions
 developed in the 70's 
\cite{Ranft:tf,Bali:as,Bali:ap,Bali:rk,Boggild:sh}
 also had this type of $dn/d\eta$. 
 As will be shortly shown, this Wood-Saxon form is not compatible with the
 transition region universality and hence 
 it is unlikely that this is the right form of $dn/d\eta$.
 Nevertheless we feel that it is worth
 considering the Wood-Saxon form here because it gives an example of slowly
 changing (as opposed to universal) limiting fragmentation curve.
 
 A reasonable description of the current data can be provided by the
 following combination of the
 Wood-Saxon (Fermi-Dirac) functions and a hyperbolic 
 cosine\cite{Eskola:2002qz}
 \be
 {dn\over d\eta}
 =
 {g\, \cosh(\eta/\zeta) 
 \over
 [1 + e^{-(\eta+\etaf)/\sigma}]
 [1 + e^{(\eta-\etaf)/\sigma}]
 }
 \label{eq:model_ws_full}
 \ee
 where $g$, $\etaf$, $\zeta$ and $\sigma$ are functions of $\ymax$. 
 Here the parameter $\etaf$ roughly corresponds to where the fragmentation
 region begins.  The hyperbolic cosine is there to provide the dip in the
 middle. Since the dip is usually shallow, $\zeta \gg \sigma$. 

 Since we are not so much interested in the dip, we consider a simplified
 form\footnote{Entirely analogous anaylsis can be also performed using 
 Eq.(\ref{eq:model_ws_full}).}
 \be
 {dn\over d\eta}
 =
 {g
 \over
 1 + e^{(\eta-\etaf)/\sigma}
 } \ \ \ \ \ \hbox{for}\ \ \eta > 0
 \label{eq:model_ws}
 \ee
 Universal fragmentation behavior demands that for high enough
 energy
 \be
 \left. {dn\over d\eta} \right|_{\eta = \ymax + \eta'} = f_U(\eta')
 \ee
 where $f_U(\eta')$ is indepdent of $\ymax$.
 If $\etaf/\sigma \gg 1$, this just implies that 
 $g = \kappa_0 e^{(\ymax - \etaf)/\sigma}$ so that it compensates the
 large exponential in the denominator.
 Near $\eta=\ymax$ this yields
 $f_U(\eta') = \kappa_0\, e^{-\eta'/\sigma}$.
 However at SPS and RHIC energy, $\ymax$ is only about 3 to 5 and it can be
 easily shown that simply having 
 $g = \kappa_0 e^{(\ymax - \etaf)/\sigma}$ with a constant $\sigma$
 does not result in the universal limiting fragmentation. 
 Instead, we must regard all parameters appearing in
 Eq.(\ref{eq:model_ws}) as functions of $\ymax$ and look for a relationship
 among them by requiring
 \be
 {\partial\over \partial\ymax} {dn\over d\eta'} 
 \approx 0
 \label{eq:ws_cond}
 \ee
 near $\eta=\ymax$ (or $\eta' = \eta-\ymax = 0)$.

 The solution of Eq.(\ref{eq:ws_cond}) is obtained as follows.
 Set $\sigma = 1/\tau$ and $\eta = \ymax + \eta'$ and $\chif = \ymax-\etaf$
 and write
 \be
 f \equiv {dn\over d\eta'} = {g\over 1 + e^{(\eta'+\chif)\tau}}
 \ee
 Taking the derivative with respect to $y = \ymax$ yields
 \be
 {\partial f\over \partial y}
 & = &
 {(dg/dy)(1+ e^{(\eta'+\chif)\tau}) 
  - g\left[ \eta'(d\tau/dy) + d(\chif\tau)/dy \right]e^{(\eta'+\chif)\tau}
  \over
 (1 + e^{(\eta'+\chif)\tau})^2}
 \ee
 If this is to be independent of $\eta'$ for small $\eta'$,
 we must have 
 \be
 (dg/dy)\left[1 + e^{-\chif\tau}(1-\eta'\tau)\right]
  \approx g\left[ \eta'(d\tau/dy) + d(\chif\tau)/dy \right]
 \ee
 which yields the following two conditions: 
 \be
 -\tau e^{-\chif\tau} {dg\over dy} = g{d\tau\over dy}
 \ee
 and
 \be
 {dg\over dy}\left[1 + e^{-\chif\tau}\right]
 = g {d(\chif\tau)\over dy}
 \ee
 Assuming monotonic functions, we can rewrite them as
 \be
 {dg\over g}
 = -{d\tau \over \tau} e^{\chif\tau}
 \ee
 and
 \be
 {dg\over g}
 =
 {d(\chif\tau)\over 1 + e^{-\chif\tau}}
 \ee
 Solving the second equation first gives
 \be
 g = \kappa_0(1 + e^{(\ymax-\etaf)/\sigma})
 \ee
 Combine the two equations to get 
 \be
  {d(\chif\tau)\over (1+e^{-\chif\tau}) }
  =
  -{d\tau \over \tau} e^{\chif\tau}
 \ee
 
 Let $L = e^{-\chif\tau}$ and use $dL = -d(\chif\tau)\, e^{-\chif\tau}$ to
 get
 \be
 -{dL\over 1 + L} = -{d\tau\over \tau}
 \ee
 Solving this equation, we finally get 
 \be
 \tau = C' (1 + e^{-\chif\tau})
 \ee
 which can be rearranged to yield
 \be
 \sigma = {\sigma_0\over 1 + e^{-(\ymax-\etaf)/\sigma}}
 \ee
 or
 \be
 {1\over \sigma} =
 {1\over \sigma_0}
 +
 {1\over \chif}\,W\left((\chif/\sigma_0)e^{-\chif/\sigma_0}\right) 
 \ee
 where we used $\chif=\ymax-\etaf$
 and
 $W(w)$ is the Lambert function that solves $w = x e^x$ for $x$.
 Hence given a value of $\etaf$, 
 the width and the height of the Wood-Saxon function is
 completely determined.
 
 \begin{figure}[t]
 \begin{center}
  \epsfxsize=0.7\tw
  \epsfbox{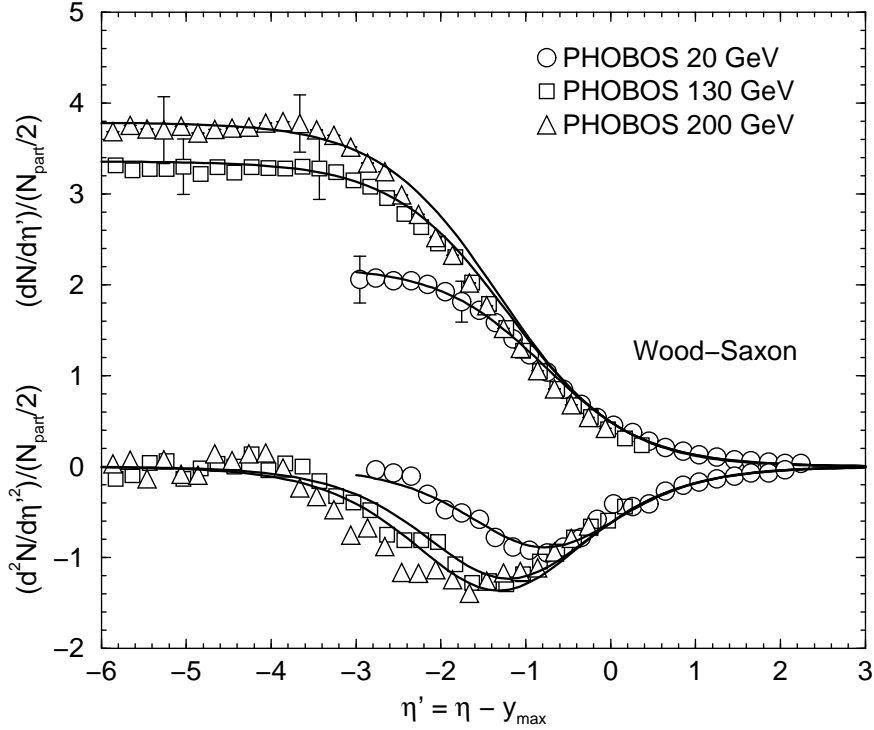}
 \end{center}
 \caption{$\{\chif,\sigma\} = \{1.32, 0.694\}$ 
 for 200\,GeV, $\{1.20, 0.681\}$ for 130\,GeV and
 $\{0.77, 0.619\}$ for 19.6\,GeV. The limiting curve parameters are
 $\kappa_0 = 0.492$ and $1/\sigma_0 = 1.253$.}
 \label{fig:wsfits}
 \end{figure}
 
 For large enough $(\ymax-\etaf)/\sigma$, 
 \be
 {1\over \sigma}
 \approx 
 {1\over \sigma_0}(1+ e^{-(\ymax-\etaf)/\sigma_0})
 \ee
 ignoring terms of $O(e^{-2(\ymax-\etaf)/\sigma_0})$.
 At an asymptotically high energy,
 \be
 \lim_{\ymax\to \infty} \sigma  = \sigma_0 
 \ee
 Hence the limiting curve is given by
 \be
 f_U(\eta') = \kappa_0\, e^{-\eta'/\sigma_0}
 \ee
 Fitting the $\eta > \ymax$ portion of $\sqrts = 19.6\,\GeV$ data yields
 $1/\sigma_0 = 1.253$ and $\kappa_0 = 0.492$ as shown in Fig.~\ref{fig:tailfit}.
 The height of the plateau and the total multiplicity can be now easily
 obtained from Eq.(\ref{eq:model_ws})
 \be
 \left(dn\over d\eta\right)_0
 &=&
 {\kappa_0 (1+e^{(\ymax-\etaf)/\sigma})\over 1 + e^{-\etaf/\sigma}}
 \label{eq:height_ws}
 \ee
 and
 \be
 n_{\rm total} 
 =
 2 \etaf\,
 \left({dn\over d\eta}\right)_{0} + O(e^{-\etaf/\sigma})
 \ee
 
 The resulting $dn/d\eta$ and $d^2n/d\eta^2$ are
 shown in Fig.~\ref{fig:wsfits} together with PHOBOS data.
 Note that although fragmentation region universality is reasonably well
 described by the Wood-Saxon functions, the transition region universality 
 is not.
 Again, we emphasize that it is the slope ($d^2n/d\eta^2$) that gives
 clearer criterion for the goodness of the description.

 Unlike the previous case, the Wood-Saxon case has no separate
 $\etap$.  This is because both the fragmentation and the transition behaves
 like an exponential $\sim e^{-|\eta-\etaf|/\sigma_0}$ near $\etaf$. 
 Therefore the transition between the plateau and the limiting
 behavior happens within about $3 \sigma_0$ around $\etaf$.
 This fact also indicates that dynamics of the plateau and the dynamics of
 the fragmenation region does {\em not} decouple even at an asymptotically
 high $\sqrts$. 
 
 This non-decoupling also allows us to put a severe condition 
 on the transverse energy.
 To calculate the energy content of the plateau, we need to carry out an
 integral over the product of the Wood-Saxon and a hyperbolic cosine.  This
 can be done, but the resulting form is not particularly illuminating.
 However, within the plateau we can approximate
 \be
 E_{\rm plateau}
 & \approx &
 2 \left(dn\over d\eta\right)_0
 \ave{m_T}_{\rm pl.} \, \sinh(\etaf)
 \ee
  Energy conservation demands that
 \be
 \ave{m_T}_{\rm pl.} < {m_N\over \kappa_0}\, e^{-(1/\sigma-1)(\ymax-\etaf)}
 \ee
 where we used Eq.(\ref{eq:height_ws}). 
 Since $\sigma < 1$, this indicates that the average transverse energy 
 in the plateau region
 must be a {\em decreasing} function of $\ymax$ if $\ymax-\etaf$ is an
 increasing function of $\ymax$.  This is an absurd result.  One would
 expect that as $\sqrts$ becomes larger,
 $\ave{m_T}_{\rm pl.}$ would also become larger or
 at least reach a limiting value, but not decrease.
 This, in our opinion, invalidates the Wood-Saxon description of $dn/d\eta$.

\begin{figure}[t]
 \begin{center}
  \epsfxsize=0.7\tw
  \epsfbox{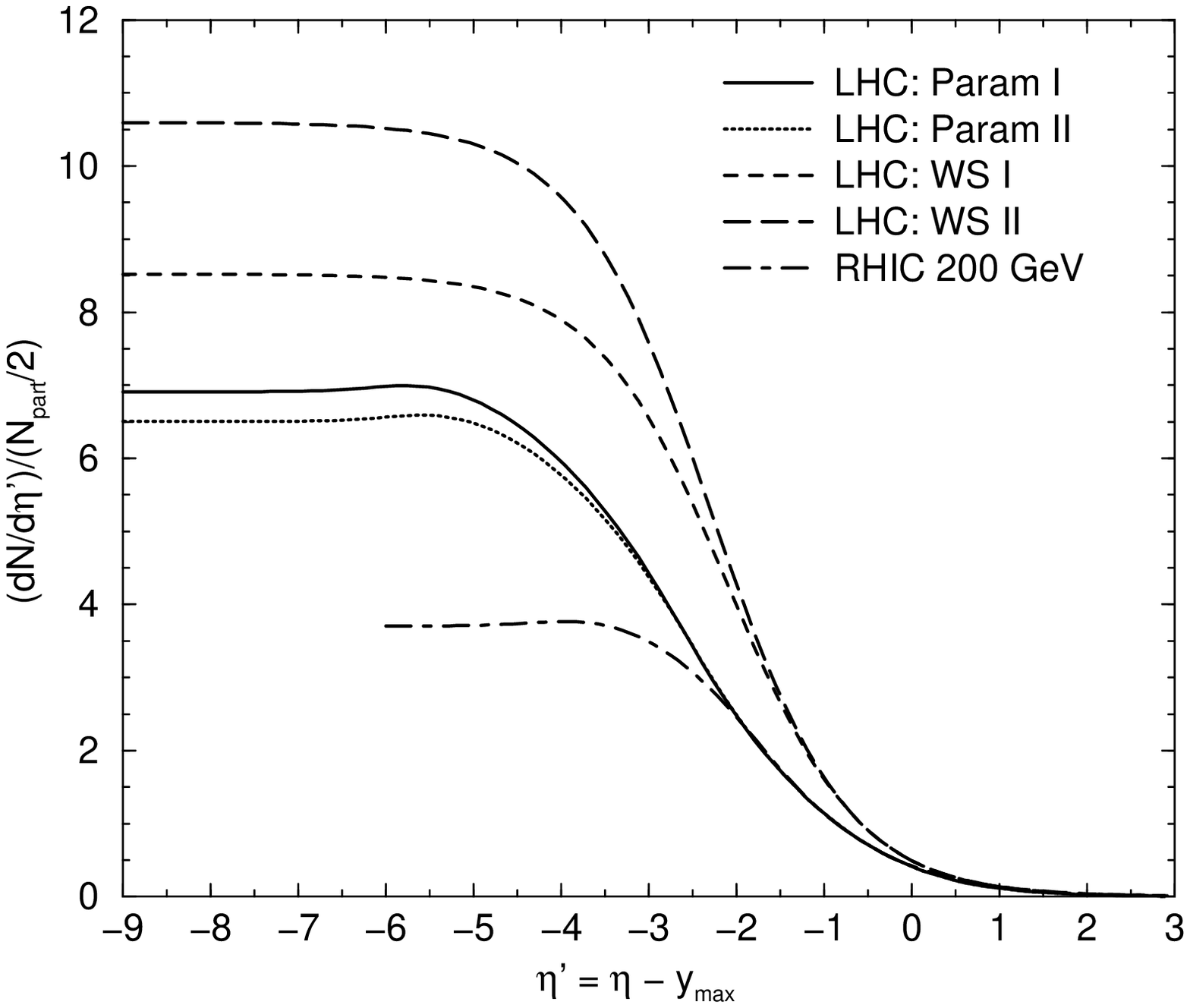}
 \end{center}
 \caption{Predictions for central collisions at LHC.  
 Also shown are parametrized RHIC 200 result.
 We set $\sqrt{s} = 5.5\,\hbox{TeV}$ and Pb+Pb for LHC.} 
 \label{fig:lhc_ws}
 \end{figure}

Nevertheless, it is instructive to also have the extrapolated Wood-Saxon
result to LHC.
For the Wood-Saxon form, we find that
$\chif = 1.32, 1.20, 0.77$ for $\sqrts = 200\,\GeV$, $130\,\GeV$,
$20\,\GeV$, respectively.  These yield 
\be
\ymax - \etaf = \left\{
\begin{array}{ll}
0.59\, \Ymax^{0.85} - 0.68\, \ln \ymax
& \ \  (\hbox{WS I})
\\
-0.47 + 0.57\, \Ymax^{0.69}
& \ \ (\hbox{WS II})
\end{array}
\right.
\ee
The results for LHC are tabulated in Table~\ref{table:ws_models}.
The values of $(dn/d\eta)_0$ and $n_{\rm total}$ are comparable to 
the saturation model (K \& L) values in Table~\ref{table:models}.
Also one can easily see in Fig.~\ref{fig:lhc_ws} that the limiting
fragmentation curve followed by the Wood-Saxon functions is not the same as
the one followed by the interpolating-exponential ones. 

\begin{table}[t]
 \begin{tabular}{|c|c|c|c|}
 \hline
 & $\ymax-\etaf$ & $(dn/d\eta)_0$ & $n_{\rm total}$ \\
 \hline
 WS I  & 2.3 & 10.6 & 130\\
 \hline
 WS II  & 2.1 & 8.5 & 110\\
 \hline
 \end{tabular}
 \caption{Predictions for LHC 
 central collisions using the Wood-Saxon form.
 We set $\sqrt{s} = 5.5\,\hbox{TeV}$ and Pb+Pb.} 
\label{table:ws_models}
\end{table}

\begin{thebibliography}{99}



\bibitem{Back:2002wb}
B.~B.~Back {\it et al.},
Phys.\ Rev.\ Lett.\  {\bf 91}, 052303 (2003) 


\bibitem{Bearden:2001qq}
I.~G.~Bearden {\it et al.}  [BRAHMS Collaboration],
Phys.\ Rev.\ Lett.\  {\bf 88}, 202301 (2002)

\bibitem{Benecke:sh}
J.~Benecke, T.~T.~Chou, C.~N.~Yang and E.~Yen,
Phys.\ Rev.\  {\bf 188}, 2159 (1969).

\bibitem{Feynman:ej}
R.~P.~Feynman,
Phys.\ Rev.\ Lett.\  {\bf 23}, 1415 (1969).

\bibitem{Hagedorn:gh}
R.~Hagedorn,
Nucl.\ Phys.\ B {\bf 24}, 93 (1970).

\bibitem{Chou:bj}
T.~T.~Chou and C.~N.~Yang,
Phys.\ Rev.\ Lett.\  {\bf 25}, 1072 (1970).

\bibitem{Chou:dh}
T.~T.~Chou and C.~N.~Yang,
Phys.\ Rev.\ D {\bf 50}, 590 (1994).


\bibitem{Dao:gg}
F.~T.~Dao {\it et al.},
Phys.\ Rev.\ Lett.\  {\bf 33}, 389 (1974).


\bibitem{Carazza}
B.~Carazza and G.~Marchesini
Phys.\ Lett.\ B {\bf 35}, 436 (1971).

\bibitem{VanderVelde:eh}
J.~C.~Vander Velde,
Phys.\ Lett.\ B {\bf 32}, 501 (1970).

\bibitem{Kita:1979ju}
I.~Kita, R.~Nakamura, R.~Nakajima and I.~Yotsuyanagi,
Prog.\ Theor.\ Phys.\  {\bf 63}, 919 (1980).

\bibitem{Mori:ay}
K.~Mori, K.~Mizutani and A.~Ogawa,
Mod.\ Phys.\ Lett.\ A {\bf 2}, 783 (1987).

\bibitem{Nakamura:1987vd}
E.~R.~Nakamura and K.~Kudo,
Z.\ Phys.\ C {\bf 40}, 81 (1988).

\bibitem{Hoang:qz}
T.~F.~Hoang,
Z.\ Phys.\ C {\bf 62}, 481 (1994).

\bibitem{Hoang:1995mk}
T.~F.~Hoang,
Z.\ Phys.\ C {\bf 68}, 467 (1995).

\bibitem{Back:2001ae}
B.~B.~Back {\it et al.}  [PHOBOS Collaboration],
Phys.\ Rev.\ Lett.\  {\bf 88}, 022302 (2002)


\bibitem{Back:2003xk}
B.~B.~Back {\it et al.}  [PHOBOS Collaboration],
arXiv:nucl-ex/0301017.

\bibitem{Jalilian-Marian:2002wq}
J.~Jalilian-Marian,
arXiv:nucl-th/0212018.

\bibitem{Iancu:2002vu}
For instance, see
E.~Iancu,
Nucl.\ Phys.\ A {\bf 715}, 219 (2003)

\bibitem{Wang:1996yf}
X.~N.~Wang,
Phys.\ Rept.\  {\bf 280}, 287 (1997)

\bibitem{Bass:1998ca}
S.~A.~Bass {\it et al.},
Prog.\ Part.\ Nucl.\ Phys.\  {\bf 41}, 225 (1998)

\bibitem{Bleicher:1999xi}
M.~Bleicher {\it et al.},
J.\ Phys.\ G {\bf 25}, 1859 (1999)


\bibitem{Kharzeev:2000ph}
D.~Kharzeev and M.~Nardi,
Phys.\ Lett.\ B {\bf 507}, 121 (2001)

\bibitem{Kharzeev:2001gp}
D.~Kharzeev and E.~Levin,
Phys.\ Lett.\ B {\bf 523}, 79 (2001)



\bibitem{Bleicher:1999pu}
M.~J.~Bleicher {\it et al.},
Phys.\ Rev.\ C {\bf 62}, 024904 (2000)

\bibitem{Hagiwara:fs}
K.~Hagiwara {\it et al.}  [Particle Data Group Collaboration],
Phys.\ Rev.\ D {\bf 66}, 010001 (2002).
See also references therein.


\bibitem{Kharzeev:2002pc}
D.~Kharzeev, E.~Levin and L.~McLerran,
arXiv:hep-ph/0210332.

\bibitem{Dumitru:2002wd}
A.~Dumitru, L.~Gerland and M.~Strikman,
Phys.\ Rev.\ Lett.\  {\bf 90}, 092301 (2003)

\bibitem{Kovchegov:1999ep}
Y.~V.~Kovchegov, E.~Levin and L.~D.~McLerran,
Phys.\ Rev.\ C {\bf 63}, 024903 (2001)

\bibitem{Abe:1989td}
F.~Abe {\it et al.}  [CDF Collaboration],
Phys.\ Rev.\ D {\bf 41}, 2330 (1990).

\bibitem{Gazdzicki:ih}
M.~Gazdzicki and O.~Hansen,
Nucl.\ Phys.\ A {\bf 528}, 754 (1991).


\bibitem{ToporPop:2002gf}
V. Topor Pop, M. Gylassy, J. Barrette, C. Gale,
X. N. Wang, N. Xu, K. Filimonov,
Phys.\ Rev.\ C {\bf 68}, 054902 (2003)



\bibitem{Wang:2000bf}
X.~N.~Wang and M.~Gyulassy,
Phys.\ Rev.\ Lett.\  {\bf 86}, 3496 (2001)

\bibitem{Li:2001xa}
S.~y.~Li and X.~N.~Wang,
Phys.\ Lett.\ B {\bf 527}, 85 (2002)

\bibitem{Back:2003hx}
B.~B.~Back {\it et al.}  [PHOBOS Collaboration],
 arXiv:nucl-ex/0311009.




\bibitem{Ranft:tf}
J.~Ranft,
Phys.\ Lett.\ B {\bf 36}, 225 (1971).

\bibitem{Bali:as}
N.~F.~Bali, L.~S.~Brown, R.~D.~Peccei and A.~Pignotti,
Phys.\ Lett.\ B {\bf 33}, 175 (1970).

\bibitem{Bali:ap}
N.~F.~Bali, L.~S.~Brown, R.~D.~Peccei and A.~Pignotti,
Phys.\ Rev.\ Lett.\  {\bf 25}, 557 (1970).

\bibitem{Bali:rk}
N.~F.~Bali, L.~S.~Brown and R.~D.~Peccei,
Phys.\ Rev.\ D {\bf 4}, 2760 (1971).

\bibitem{Boggild:sh}
H.~Boggild, K.~H.~Hansen and M.~Suk,
Nucl.\ Phys.\ B {\bf 27}, 1 (1971).

\bibitem{Eskola:2002qz}
K.~J.~Eskola, K.~Kajantie, P.~V.~Ruuskanen and K.~Tuominen,
Phys.\ Lett.\ B {\bf 543}, 208 (2002)

\bibitem{integ:GR}
I.~S.~Gradshteyn and I.~M.~Ryzhik,
``Table of Integrals, Series, and Products'', Academic Press Inc., San
Diego, CA, (1980).
 
\bibitem{phobosweb} http://www.phobos.bnl.gov

\end{thebibliography}
\end{document}